\documentclass[twocolumn,pre,aps,floatfix,longbibliography,superscriptaddress]{revtex4-1} 
\usepackage{color}
\usepackage{graphicx}
\usepackage[color=green!60]{todonotes}
\usepackage{physics}
\usepackage{amsthm}
\usepackage{amsmath}
\usepackage{amssymb}
\usepackage{enumerate}
\usepackage{placeins}
\usepackage{booktabs}
\usepackage{dsfont}
\usepackage{hyperref}
\usepackage{multirow}
\usepackage[T1]{fontenc}
\usepackage{xcolor}
\usepackage{mathtools}

\newcommand{\multiset}[2]{\ensuremath{\left(\kern-.3em\left(\genfrac{}{}{0pt}{}{#1}{#2}\right)\kern-.3em\right)}} 

\renewcommand{\S}{Sec.}
\renewcommand{\vec}[1]{\boldsymbol{#1}}
\newcommand{\e}[1]{\mathrm{e}^{-\beta #1}}
\newcommand{\ep}[1]{\mathrm{e}^{\beta #1}}
\newcommand{\en}[2]{\mathrm{e}^{-#1\beta #2}}

\newcommand{\Z}[2]{Z_{\!{\scriptscriptstyle#1}{\!,#2}}}
\newcommand{\Y}[2]{Y_{\!{\scriptscriptstyle#1}{\!,#2}}}
\newcommand{\ZZ}[3]{Z_{\!{\scriptscriptstyle#1}{\!,#2}}^{\setminus \{{#3}\}}}
\newcommand{\ZZZ}[3]{Z_{\!{\scriptscriptstyle#1}{\!,#2}}^{\cup \qty{{#3}}}}
\newcommand{\ZZs}[3]{Z_{\!{\scriptscriptstyle#1}{\!,#2}}^{\setminus {#3}}}
\newcommand{\ZZZs}[3]{Z_{\!{\scriptscriptstyle#1}{\!,#2}}^{\cup {#3}}}
\newcommand{\jd}[1]{j^{\scriptscriptstyle(#1)}}
\newcommand{\jdd}[2]{j_{#1}^{\scriptscriptstyle(#2)}}
\newcommand{\p}[2]{\mathcal{P}_{\!\!{\scriptscriptstyle#1}{\!,#2}}}

\newcommand{\avg}[3]{\expval{#3}_{\!{\scriptscriptstyle#1}{\!,#2}}}
\newcommand{\avgsq}[3]{\expval{#3}^2_{\!{\scriptscriptstyle#1}{\!,#2}}}
\newcommand{\subj}[1]{\vert_{{{#1}}}}

\newcommand{\Eqref}[1]{Eq.~\eqref{#1}}
\AtBeginDocument{%
\heavyrulewidth=.08em
\lightrulewidth=.05em
\cmidrulewidth=.03em
\belowrulesep=.65ex
\belowbottomsep=0pt
\aboverulesep=.4ex
\abovetopsep=0pt
\cmidrulesep=\doublerulesep
\cmidrulekern=.5em
\defaultaddspace=.5em
}

\begin{document}

\title{Theory of Non-Interacting Fermions and Bosons in the Canonical Ensemble}

\author{Hatem Barghathi}
% \ead{Hatem.Barghathi@uvm.edu}
\affiliation{Department of Physics, University of Vermont, Burlington, VT 05405, USA}
\affiliation{Department of Physics, Missouri University of Science and Technology, Rolla, MO 65409, USA}

\author{Jiangyong Yu}
\affiliation{Department of Physics, University of Vermont, Burlington, VT 05405, USA}

\author{Adrian Del Maestro}
\affiliation{Department of Physics and Astronomy, University of Tennessee, Knoxville, TN 37996, USA}
\affiliation{Min H. Kao Department of Electrical Engineering and Computer Science, University of Tennessee, Knoxville, TN 37996, USA}
\affiliation{Department of Physics, University of Vermont, Burlington, VT 05405, USA}

\begin{abstract}
We present a self-contained theory for the exact calculation of particle number counting statistics of non-interacting indistinguishable particles in the canonical ensemble. This general framework introduces the concept of auxiliary partition functions, and represents a unification of previous distinct approaches with many known results appearing as direct consequences of the developed mathematical structure.  In addition, we introduce a general decomposition of the correlations between occupation numbers in terms of the occupation numbers of individual energy levels, that is valid for both non-degenerate and degenerate spectra.  To demonstrate the applicability of the theory in the presence of degeneracy, we compute energy level correlations up to fourth order in a bosonic ring in the presence of a magnetic field.
\end{abstract}

\maketitle

% ---------------------------------------------------------------------------------
% Introduction
\section{Introduction}

In quantum statistical physics, the analysis of a fixed number $N$ of indistinguishable particles is difficult, even in the non-interacting limit. In such a canonical ensemble, the constraint on the total number of particles gives rise to correlations between the occupation probability and the corresponding occupation numbers of the available energy levels. As a result, the canonical treatment of non-interacting fermions and bosons is in general avoided,  and it is often completely absent in introductory texts addressing quantum statistical physics \cite{kittel1980thermal,LL5,pathria2011statistical}.  The standard approach is to relax the fixed $N$ constraint and instead describe the system using the grand canonical ensemble. While in most cases this approximation can be trusted in the large$-N$ limit, it can explicitly fail in the low-temperature regime  \cite{Bedingham:2003,MullinFernandez:2003}. The situation is even worse if the low-temperature system under study is mesoscopic, containing a relatively small number of particles.  

Recent advances in ultra-cold atom experiments \cite{Wenz_etal:2013,Parsons_etal:2015,Cheuk_etal:2015,Haller_etal:2015,PegahanKangaraArakelyanThomas_2019,Mukherjee_etal:2017,Hueck_etal:2018,Mukherjee_etal:2019,Onofrio:2016,Phelps_etal:2020} are making such conditions the rule rather than the exception, where the interactions between particles are often ignored, especially in the fermionic case  \cite{Giorgini_etal:2008}. Thus an accurate statistical representation of these systems should be canonical, enforcing the existence of fixed $N$. 
In a different context, particle number conservation can reduce the amount of quantum information (entanglement) that can be extracted from a quantum state \cite{Horodecki:2000,BartlettWiseman:2003,WisemanVaccaro:2003,WisemanBartlettVaccaro:2004,VaccaroAnselmiWiseman:2003,SchuchVerstraeteCirac:2004,DunninghamRauBurnett:2005,CramerPlenioWunderlich:2011,KlichLevitov:2008} and thus the fixed $N$ constraint requires a canonical treatment. This is reflected in the calculation of the symmetry resolved entanglement which has recently been studied in a variety of physical systems \cite{MurcianoRuggieroCalabrese:2020,TanRyu:2020,MurcianoDiGiulioCalabrese:2020arXiv,CapizziRuggieroCalabrese:2020,FraenkelGoldstein:2020,FeldmanGoldstein:2019,BarghathiCasiano-DiazDelMaestro:2019,BonsignoriRuggieroCalabrese:2019,BarghathiHerdmanDelMaestro:2018,KieferEmmanouilidisUnanyanFleischhauer:2020,MurcianoDiGiulioCalabrese:2020,GoldsteinSela:2018}. In more general settings,  the presence of conservation laws could demand a canonical treatment, as in the case of nuclear statistical models \cite{ Sato:1987,ChengPratt:2003,PrattRuppert:2003,ToneevParvan:2005,AkkelinSinyukov:2016,ParvanToneevPloszajczak:2000,Das_etal:2005,JenningsDasGupta:2000,Rossignoli:1995,CanosaRossignoliRing:1999,RossignoliCanosaEgido:1996,Gudima_etal:2000,Vovchenko_etal:2019,Vovchenko_etal:2018,JiaQi:2016,BegunGorensteinZozulya:2005,Jing-Hua:2017,Garg_etal:2016,Acharya_etal:2020}. 

Given its pedagogical and now practical importance, as well as the long history of the problem of non-interacting quantum particles, the last 50 years has provided a host of results, varying from general recursive relations that govern the canonical partition functions and the corresponding occupation numbers \cite{Schmidt:1989,BorrmannFranke:1993,Borrmann_etal:1999,Pratt:2000,MullinFernandez:2003,WeissWilkens:1997,Arnaud_etal:1999,Schonhammer:2017,GiraudGrabschTexier:2018,TsutsuiKita:2016,ZhouDai:2018,Dean_etal:2016}, to approximate \cite{DentonMuhlschlegelScalapino:1971,DentonMuhlschlegelScalapino:1973, SchonhammerMeden:1996,Bedingham:2003,Svidzinsky_etal:2018,JhaHirata:2020} and exact results for some special cases \cite{ChaseMekjianZamick:1999,Schonhammer:2000,Kocharovsky_etal:2010,WangMa:2009,MagnusLemmensBrosens:2017,Grabsch_etal:2018,Kruk_etal:2020,GrelaMajumdarSchehr:2017,LiechtyWang:2020}.  More recently, for the case of non-degenerate energy spectra, the exact decomposition of higher-order occupation number correlations in terms of the occupation numbers of individual bosonic and fermionic energy levels have been reported~\cite{Schonhammer:2017,GiraudGrabschTexier:2018}.  

In this paper, we present a unified framework for the calculation of physical observables in the canonical ensemble for $N$ fermions or bosons that are applicable to general non-interacting Hamiltonians that may contain degenerate energy levels.  We present an analysis of the mathematical structure of bosonic and fermionic canonical partition functions in the non-interacting limit that leads to a set of recursion relations for exactly calculating the energy level occupation probabilities and average occupation numbers.  Using argument from linear algebra, we show how higher-order correlations between the occupation numbers can be factorized, allowing them to be obtained from the knowledge of the occupation numbers of the corresponding energy levels, a canonical generalization of Wick's theorem \cite{Wick:1950hn}. The key observation yielding simplification of calculations in the canonical ensemble is that occupation probabilities and occupation numbers can be expressed via auxiliary partition functions (APFs) -- canonical fermionic or bosonic partition functions that correspond to a set of energy levels that are obtained from the full spectrum of the targeted system by making a subset of the energy levels degenerate (increasing its degeneracy if its already degenerate) or alternatively by excluding it from the spectrum. Results obtained via auxiliary partition functions are validated by demonstrating that a number of previously reported formulas can be naturally recovered in a straightforward fashion within this framework.

In a non-interacting system, observables such as the average energy and magnetization can be obtained solely from the knowledge of average occupation numbers, however, the calculation of the corresponding statistical fluctuations in such quantities, i.e., specific heat and magnetic susceptibility, requires knowledge of the fluctuations in occupation numbers and (in the canonical ensemble) correlations between them. Therefore, the factorization of correlations between occupation numbers in terms of the average occupation numbers of individual energy levels provides a simplified approach to calculate quantities such as specific heat and magnetic susceptibility of a given system. Here, we highlight a method to compute such correlations by considering a degenerate system of a finite periodic chain of non-interacting bosons that is influenced by an external uniform magnetic field.

The auxiliary partition function approach presented here provides a set of tools that can be used to analyze experimental data in low-density atomic gases where the number of particles is fixed.  The application of the physically relevant canonical ensemble can eliminate errors introduced via the grand canonical approximation (especially in the inferred temperature) and lead to a more accurate interpretation of experimental results, including an improved diagnosis of the role of weak interaction effects. 
It is hoped that the relative simplicity of the mathematical approach presented in this paper may encourage the inclusion of the interesting topic of the canonical treatment of Fermi and Bose gases in college-level textbooks.

In \S\ref{sec:II} we present the general formalism of the theory, where we introduce APFs and write occupation probability distributions, occupation numbers, and their correlations in terms of the APFs. In the same section, we also show how some of the previously known results could be recovered in a straightforward fashion. In \S\ref{sec:III}, considering fermionic and bosonic systems, we derive the decomposition of level higher-order occupation numbers correlations into individual energy levels occupation numbers for non-degenerate and degenerate energy spectra, alike. To illustrate the applicability of the theory, In \S\ref{sec:IV},  we consider a bosonic ring in the presence of a magnetic field. We conclude in \S\ref{sec:V}.

% ---------------------------------------------------------------------------------
\section{Non-interacting Indistinguishable particles in the canonical ensemble}
\label{sec:II}
% ---------------------------------------------------------------------------------
As the thermodynamic properties of a system of non-interacting particles are governed by the single-particle spectrum and the underlying particle statistics, we begin by considering the general one-particle spectrum $\epsilon_i$, with $i\in\mathcal{S}=\{1, 2, ,\dots, M\}$. For an unbounded spectrum $M\to\infty$. In the  canonical ensemble defined by fixing the total particle number $N$, the canonical partition function $Z_{N}\equiv Z_{N}(\mathcal{S})$ for $N$ indistinguishable particles is defined by:
%fermions and bosons
\begin{equation}
    Z_{N}=\sum_{\vec{n}\subj{N}}X\qty(\vec{n}\subj{N})
\label{eq:ZN},
\end{equation}
where  
\begin{equation}
 X(\vec{n}\vert_{N})=\prod_{i\in \mathcal{S}} \e{\epsilon_in_i} 
\label{eq:Xboltzmann}
\end{equation}
are the Boltzmann factors at inverse temperature $\beta = 1/k_{\rm B} T$ and the components $\{n_i\}$ of the vector $\vec{n}\vert_{N} = (n_1,\dots, n_M )\vert_N $ are the occupation numbers for the corresponding energy levels satisfying $\sum_{i\in\mathcal{S}} n_i=N$. The summation in Eq.~(\ref{eq:ZN}) runs over all the possible occupation vectors $\vec{n}\vert_{N}$ which, in addition to conserving the total number of particles $N$, obeys occupation limits for each of the energy levels $\epsilon_i$: $n_i \le n_i^{\max}$, thus $N\le N_{\max}=\sum_i n_i^{\max}$.  
%We use the convention that $Z_{0}=1$. 

Consider the spectrum defined by $\mathcal{S}$ as the union of disjoint subsets (subspectra) $\mathcal{S}^{(1)}$ and $\mathcal{S}^{(2)} = \mathcal{S} \setminus S^{(1)}$, i.e., $\mathcal{S}=\mathcal{S}^{(1)} \cup \mathcal{S}^{(2)}$. Under this decomposition the Boltzmann factors in $\mathcal{S}$ can be factorized as: $X\qty(\vec{n}\vert_{N})=X\qty(\vec{n}^{(1)}\vert_{k})X\qty(\vec{n}^{(2)}\vert_{N-k})$, where $\vec{n}^{(1)}\vert_{k}$ and $\vec{n}^{(2)}\vert_{N-k}$ represent occupation vectors of $k = \sum_{i \in \mathcal{S}^{(1)}}n_i^{(1)}$ and $N-k$ particles in $\mathcal{S}^{(1)}$ and $\mathcal{S}^{(2)}$, respectively. Summing $X\qty(\vec{n}\vert_{N})$ over all possible $\vec{n}^{(1)}\vert_{k}$ and $\vec{n}^{(2)}\vert_{N-k}$ gives: 
\begin{equation}
\sum_{\vec{n}^{(1)}\vert_{k}}\sum_{\vec{n}^{(2)}\vert_{N-k}}X\qty(\vec{n}\vert_{N})=Z_k\qty(\mathcal{S}^{(1)}) Z_{N-k}\qty(\mathcal{S}^{(2)}),  
\label{eq:ZS1S2}
\end{equation}
where we have introduced the APFs:
\begin{align}
\label{eq:auxZk}
Z_k\qty(\mathcal{S}^{(1)}) &=\sum_{\vec{n}^{(1)}\vert_{k}}X\qty(\vec{n}^{(1)}\vert_{k}) \\
 Z_{N-k}\qty(\mathcal{S}^{(2)}) &=\sum_{\vec{n}^{(2)}\vert_{N-k}}X\qty(\vec{n}^{(2)}\vert_{N-k})\,.
\label{eq:auxZNmk}
\end{align}
For Eqs.~(\ref{eq:auxZk}) and (\ref{eq:auxZNmk}) to satisfy the restrictions imposed by the per-energy level maximum occupancies $\{n_i^{\max}\}$ and the fixed $N$, $k$ must satisfy $\max(0,N-N^{\max}_2)\le k \le \min(N,N^{\max}_1)$, where $N^{\max}_1$ and $N^{\max}_2$ are the maximum numbers of particles allowed in $\mathcal{S}^{(1)}$ and $\mathcal{S}^{(2)}$, respectively. Additionally, any vector $\vec{n}\vert_{N}$, under these constraints can be decomposed into two allowed vectors $\vec{n}^{(1)}\vert_{k}$ and $\vec{n}^{(2)}\vert_{N-k}$ and vice versa. Thus the sum in Eq.~(\ref{eq:ZN}) can be similarly decomposed as: $\sum_{\vec{n}\vert_{N}}\equiv \sum_{k=k_{\min}}^{k_{\max}}\sum_{\vec{n}^{(1)}\vert_{k}}\sum_{\vec{n}^{(2)}\vert_{N-k}}$ where $k_{\min} = \max (0, N-N^{(2)}_{\max})$ and $k_{\max} = \min (N, N^{(1)}_{\max})$ yielding the full partition function
\begin{equation}
Z_{N}\equiv\sum_{k=k_{\min}}^{k_{\max}}Z_k\qty(\mathcal{S}^{(1)}) Z_{N-k}\qty(\mathcal{S}^{(2)})\,.
\label{eq:Z1Z2}
\end{equation}

Employing the convention that $Z_{N}(\mathcal{S})=0$ whenever $N$ is negative, or when it exceeds the maximum number of particles set by $\mathcal{S}$, the limits in the above summation can be simplified to $\sum_{k=0}^N$. The above notation can be made more explicit by specifying the subset of levels that are not included in the partition function
\begin{equation}
    Z_{N}^{\setminus\mathcal{S}^{(1)}} \equiv Z_{N}\qty(\mathcal{S}\setminus\mathcal{S}^{(1)}) 
\label{eq:ZNsetminus}
\end{equation}
and thus for $\mathcal{S}^{(1)}=\{j_1, j_2, \dots, j_\ell\}$ containing $\ell$ levels, Eq.~(\ref{eq:Z1Z2}) is equivalent to 
%$\mathcal{S}^{(1)}=\mathcal{S}_{\ell}=\{j_1, j_2, \dots, j_\ell\}$
%
\begin{equation}
Z_{N}\equiv\sum_{k=0}^{N}Z_k(\{j_1, j_2, \dots, j_\ell\}) Z_{N-k}^{\setminus\{j_1, j_2, \dots, j_\ell\}}
\label{eq:Z1Z1}.
\end{equation}

To obtain a physical interpretation of Eq.~(\ref{eq:Z1Z1}), recall that in the canonical ensemble, the likelihood of the $N$-particle system being in a microstate defined by the occupation vector $\vec{n}\vert_{N}$ is given by the ratio $X(\vec{n}\vert_{N})/Z_N$. Accordingly, for the subset of energy levels with indices $\{j_1, j_2, \dots, j_\ell\}$, the joint probability distribution of the corresponding occupation numbers $\mathcal{P}_{n_{j_1}, n_{j_2}, \dots, n_{j_\ell}}$ can be obtained by performing the summation $\sum_{\vec{n}^{(2)}\vert_{N-k}}$,  where $k = \sum_{r=1}^\ell n_{j_r}$, yielding 
\begin{equation}
\mathcal{P}_{n_{j_1}, n_{j_2}, \dots, n_{j_\ell}}=\frac{\e{\sum_{r=1}^\ell \epsilon_{j_r}n_{j_r}}}{Z_N}Z_{N-k}^{\setminus\{j_1, j_2, \dots, j_\ell\}}
\label{eq:pn1tol}.
\end{equation}
It follows that the probability $P_k(\{j_1, j_2, \dots, j_\ell\})$ of finding $k$ particles in $\{j_1, j_2, \dots, j_\ell\}$ and, of course, $N-k$ particles in $\mathcal{S}\setminus\{j_1, j_2, \dots, j_\ell\}$, can be obtained by applying the summation $\sum_{\vec{n}^{(1)}\vert_{k}}$:
\begin{equation}
P_k(\{j_1, j_2, \dots, j_\ell\})=\frac{Z_k(\{j_1, j_2, \dots, j_\ell\}) Z_{N-k}^{\setminus\{j_1, j_2, \dots, j_\ell\}}}{Z_N}
\label{eq:pk},
\end{equation}
where the normalization of $P_k(\{j_1, j_2, \dots, j_\ell\})$ is guaranteed by 
Eq.(\ref{eq:Z1Z1}). 

So far the analysis of the partition functions has been completely general and we have not specified what type of particles are being described. Now, let's be more specific and consider a system that solely consists of either fermions or bosons.

% ---------------------------------------------------------------------------------
\subsection{An Inverted Analogy Between Fermionic and Bosonic Statistics}
\label{subsec:inverted}

Having developed an intuition for the general structure of the canonical partition function under a bipartition into sub-spectra, we now observe how this can provide insights into the relationship between fermionic ($n_i^{\max}=1$) and bosonic ($n_i^{\max}\to\infty$)  statistics of $N$ non-interacting particles. To distinguish the two cases we introduce a new subscript on the partition function ($F$ for fermions and $B$ for bosons). 

If the set $\mathcal{S}^{(1)}$ represent a single energy level with an index $j_1=j$ then using our convention we have:
\begin{equation}
    \Z{F}{k}(\{j\}) = 
    \begin{cases} 
        \e{\epsilon_j k} & 0 \le k\le 1 \\
        0 & \text{otherwise}
        \end{cases}
\label{eq:ZjFermions}
\end{equation}
for fermions and
\begin{equation}
    \Z{B}{k}(\{j\}) = 
    \begin{cases} 
        \e{\epsilon_j k} & k\ge 0 \\
        0 & \text{otherwise}
        \end{cases}
\label{eq:ZjBosons}
\end{equation}
for bosons. Substituting into Eq.~(\ref{eq:Z1Z1}) we immediately find:
\begin{align}
\label{eq:ZNF}
    \Z{F}{N} &= \ZZ{F}{N}{j}+\e{\epsilon_j}\ZZ{F}{N-1}{j} \\
    \Z{B}{N} & =\sum_{k=0}^{N}\e{\epsilon_j k}\ZZ{B}{N-k}{j}\,.
\label{eq:ZNB}
\end{align}
The relations in Eq.~(\ref{eq:ZNF}) and (\ref{eq:ZNB}) formally describe the procedure for generating the canonical partition function of the $N$ particle system after introducing an energy level $\epsilon_j$ to the preexisting spectrum $\mathcal{S}\setminus\{j\}$.  

Examining the structure of Eq.~(\ref{eq:ZNF}) suggests a simple matrix form: $\vec{Z}= \mathsf{A}\vec{Z}^{\setminus\{j\}}$, where $\vec{Z}=(\Z{F}{0}, \Z{F}{1}, \dots)$, $\vec{Z}^{\setminus\{j\}}=(\ZZ{F}{0}{j}, \ZZ{F}{1}{j}, \dots)$ and the matrix $\mathsf{A}_{n,m}=\delta_{n,m}+\e{\epsilon_j}\delta_{n,m+1}$ is bidiagonal and can be inverted such that:
\begin{equation}
\ZZ{F}{N}{j}=\sum_{k=0}^{N}(-1)^{k}\e{\epsilon_j k}\Z F {N-k}
\label{eq:ZNF_inv}.
\end{equation}
Comparing this expression with Eq.~(\ref{eq:ZNB}), we observe an identical structure apart from exchanging the factor $\e{\epsilon_j}$ with $\qty(-\e{\epsilon_j})$. Thus we can obtain the inversion of  Eq.~(\ref{eq:ZNB}) by replacing $\e{\epsilon_j}$ with $\qty(-\e{\epsilon_j})$ in Eq.~(\ref{eq:ZNF}), i.e.,
\begin{equation}
\ZZ B N j=\Z B N-\e{\epsilon_j}\Z B {N-1}
\label{eq:ZNB_inv}.
\end{equation}
The relations  Eq.~(\ref{eq:ZNF_inv}) and Eq.~(\ref{eq:ZNB_inv}) exemplify the elimination of an energy level as they represent the inverse of Eq.~(\ref{eq:ZNF}) and Eq.~(\ref{eq:ZNB}) respectively.

If we absorb the negative signs in  Eqs.~(\ref{eq:ZNF_inv})  by shifting the energy $\epsilon_j$  by $\pm i\pi/\beta$, we can write $\ZZ F N j=\sum_{k=0}^{N}\e{\epsilon^\prime_j k}\Z F {N-k}$, where $\epsilon^\prime_j=\epsilon_j\pm i\pi/\beta$. As the general bipartition into sub-spectra introduced in Eq.~(\ref{eq:Z1Z2}) holds for energy levels with mixed statistics, and doesn't require real entries for the $\epsilon_i$, we can build the bosonic partition functions $\Z B N^\prime (\{j_1, j_2, \dots, j_\ell\}) $ using the shifted energies $\{\epsilon_{j_1}^\prime, \epsilon_{j_2}^\prime, \dots, \epsilon_{j_\ell}^\prime\}$ and then combine it with the $\Z F N$ to generate a mixed-form:
\begin{equation}
\ZZ F N {j_1, j_2, \dots, j_\ell}=\sum_{k=0}^{N}\Z B k^{\prime} (\{j_1, j_2, \dots, j_\ell\}) \Z F {N-k}\,.
\end{equation}
The effect of shifting the single particle spectrum by a constant $\omega$ on the canonical partition function $Z_N$ is captured by a rescaling factor $\e {\omega N}$ and the resulting $N$  particle partition function of the shifted spectrum is $\e {\omega N}Z_N$.  Using $\Z B k^{\prime} (\{j_1, j_2, \dots, j_\ell\})={\rm{e}}^{\pm i\pi k}\Z B k (\{j_1, j_2, \dots, j_\ell\})$ then yields
\begin{equation}
\ZZ F N {j_1, j_2, \dots, j_\ell}=\sum_{k=0}^{N}(-1)^k\Z B k (\{j_1, j_2, \dots, j_\ell\}) \Z F {N-k}
\label{eq:Z1Z1F_inv}.
\end{equation}
Starting from Eq.~(\ref{eq:ZNB_inv})  and following the same argument, we obtain an equivalent expression for bosons
\begin{equation}
\ZZ B N {j_1, j_2, \dots, j_\ell}=\sum_{k=0}^{N}(-1)^k\Z F k (\{j_1, j_2, \dots, j_\ell\}) \Z B {N-k}
\label{eq:Z1Z1B_inv}.
\end{equation}
The last two equations can be seen as a generalization of Eqs.~(\ref{eq:ZNF_inv}) and (\ref{eq:ZNB_inv}). However, they also recover the symmetry between fermionic and bosonic statistics. To this end, we show how to obtain the partition function of a given spectrum using the APFs of two complementary subsets of the spectrum through \Eqref{eq:Z1Z1}. Also, Eqs.~(\ref{eq:Z1Z1F_inv}) and (\ref{eq:Z1Z1B_inv}) show that this relation can be inverted, \emph{i.e.}, we can calculate the APFs of a subset of energy level using the APFs of its complement with the opposite statistics and the partition functions of the full spectrum.

% ---------------------------------------------------------------------------------
\subsection{Energy Level Occupations and Correlations}

For fermions, the Pauli exclusion principle restricts the number of particles occupying an energy level $\epsilon_j$ to  $n_j=0$ or $1$. Equivalently, $n_j^p=n_j$ for any $p>0$, simplifying the calculation of energy level occupation numbers and the correlations between them, including higher moments. More specifically, the average $\avg F N {n_{j_1}^{p_1}n_{j_2}^{p_2}\dots n_{j_\ell}^{p_\ell}}=\avg F N {n_{j_1}n_{j_2}\dots n_{j_\ell}}$, where $p_i>0,\ i\in\{1, \dots, \ell\}$ and in general
\begin{multline}
\avg F N {n_{j_1}n_{j_2}\dots n_{j_\ell}}= \\ \sum_{\vec{n}(\{j_1, j_2, \dots, j_\ell\})}  n_{j_1}n_{j_2}\dots n_{j_\ell}\p F{n_{j_1}, n_{j_2}, \dots, n_{j_\ell}} 
\label{eq:nj1njldefF}
\end{multline}
using the probability:
\begin{equation}
    \p F{n_{j_1}, n_{j_2}, \dots, n_{j_\ell}}=\frac{\e{\sum_{r=1}^\ell \epsilon_{j_r}n_{j_r}}}{\Z F N} \ZZ F{N-\sum_{r=1}^{\ell} n_{j_r}} {j_1, j_2, \dots, j_\ell}
\label{eq:pn1tolF},
\end{equation}
defined in Eq.~(\ref{eq:pn1tol}). The only term that survives in Eq.~(\ref{eq:nj1njldefF}) has $n_{j_{1}}=n_{j_{2}}=\dots=n_{j_{\ell}}=1$ giving: 
\begin{equation}
\avg F N {n_{j_1}n_{j_2}\dots n_{j_\ell}}=\frac{\e{\sum_{r=1}^\ell \epsilon_{j_r}}}{\Z F N}\ZZ F {N-\ell}{j_1, j_2, \dots, j_\ell}
\label{eq:corrF}.
\end{equation}
Following the same procedure, \Eqref{eq:nj1njldefF} can be generalized to describe both correlations and anti-correlations between energy levels:
\begin{multline}
    \avg{F}{N}{\prod_{r=1}^{\ell}{\qty[n_{j_r}\gamma_{j_r}+(1-n_{j_r})(1-\gamma_{j_r})]}} = \\
    \frac{1}{\Z{F}{N}} \e{\sum_{r=1}^{\ell}\epsilon_{j_r}\gamma_{j_r}}\ZZ F {N-\sum_{r=1}^{\ell}\gamma_{j_r}} 
{j_1, j_2, \dots, j_\ell}
\label{eq:corrFgen},
\end{multline}
where $\gamma_{j_r} = 1,0$.  For the latter with $\gamma_{j_r}=0$, the occupation numbers in  \Eqref{eq:nj1njldefF} have been replaced with their complements, $1-n_{j_r}$.  
Thus, we find that fermionic level occupations and correlations can be directly written in terms of APFs without resorting to the usual definition $\avg F N {n_{j_1}n_{j_2}\dots n_{j_\ell}}= \frac{1}{\Z{F}{N}}\frac{\partial^{\ell}\Z F N}{\partial (-\beta\epsilon_{j_1})\dots \partial (-\beta\epsilon_{j_\ell})}$ which yields equivalent results.

When considering a single level $(\ell=1)$, the occupation probability of the $j^{\rm th}$ fermionic level immediately follows
\begin{equation}
    \avg F N  {n_j} =\frac{\e {\epsilon_j}\ZZ F {N-1} j }{\Z F N}
\label{eq:njF},
\end{equation}
with the associated probability
\begin{equation}
\p F{n_j}=\frac{\e{\epsilon_j n_j}}{\Z F N}\ZZ F {N-n_j}{j}
\label{eq:pnF}.
\end{equation}

For bosons, the occupation numbers $\avg B N {n_j}$ can be calculated from the corresponding occupation probability distribution $\mathcal{P}_{n_j}$ which are obtained from Eq.~(\ref{eq:pn1tol}):
\begin{equation}
\p B{n_j}=\frac{\e{\epsilon_j n_j}}{\Z B N}\ZZ B {N-n_j}{j}
\label{eq:pnB},
\end{equation}
and for the $\ell$-point correlations:
\begin{equation}
    \p B{n_{j_1}, n_{j_2}, \dots, n_{j_\ell}}=\frac{\e{\sum_{r=1}^\ell \epsilon_{j_r}n_{j_r}}}{\Z B N} \ZZ B {N-\sum_{r=1}^{\ell} n_{j_r}} {j_1, j_2, \dots, j_\ell}
\label{eq:pn1tolB}.
\end{equation}
However, unlike the fermionic case, such an approach requires performing the unrestricted summation $\sum_{\vec{n}(\{j_1, j_2, \dots, j_\ell\})}$.  

An alternative method which avoids this difficulty can be developed by exploiting the inverted analogy between fermionic and bosonic statistics introduced in \S\ref{subsec:inverted}.  In the fermionic case, the occupation number of an energy level $\epsilon_j$ is proportional to the APF $\ZZ F {N-1} j$  (Eq.~(\ref{eq:njF})) which corresponds to the actual spectrum of the system missing the energy level $\epsilon_j$. 
This suggests a route forward for bosons via the analogous inversion of doubly including the energy level $\epsilon_j$ instead of removing it, \emph{i.e.}, we construct an APF where this level is twofold-degenerate. We denote the corresponding $N$-boson APF by $\ZZZ{B}{N}{j}$ and distinguish the two levels using the dressed indices $\jd{0}$ and $\jd{1}$ such that the resulting combined spectrum has level indices $\qty{1,\dots j-1,\jd{0},\jd{1},j+1,\dots,M}$ where $\epsilon_j = \epsilon_{\jd{0}}= \epsilon_{\jd{1}}$. 

Returning to the general definition of the canonical partition function in Eq.~(\ref{eq:ZN}), we can write $\ZZZ B{N}{j}=\sum_{\vec{n}^{\prime}\vert_{N}} X(\vec{n}^{\prime}\vert_{N})$, where the occupation vectors $\vec{n}^{\prime}\vert_{N}$ have one extra component: $n_j$ in $\vec{n}\vert_{N}$ is replaced by $n_{\jd{0}}$ and $n_{\jd{1}}$. The modified Boltzmann factors are:
\begin{equation}
    X(\vec{n}^{\prime}\vert_{N})=\e{\epsilon_j [n_{\jd{0}}+n_{\jd{1}}]}\prod_{i\neq \jd{0}, \jd{1}} \e{\epsilon_in_i}
\label{eq:Xdoubly}
\end{equation}
and thus their value is dependent only on the total occupancy of the $j^{\rm th}$ level, $n_{\jd{0}} + n_{\jd{1}}$.  As a result, $X(\vec{n}\vert_{N})=X(\vec{n}^{\prime}\vert_{N})$ for any occupation vectors $\vec{n}\vert_{N}$ and $\vec{n}^{\prime}\vert_{N}$ with all $i\ne j$ components equal as well as $[\vec{n}\vert_N]_j = n_j=n_{\jd{0}}+n_{\jd{1}}$.  For fixed $\vec{n}\vert_{N}$, the number of vectors $\vec{n}^{\prime}\vert_{N}$ that satisfies the previous conditions is equal to $n_j+1$, or, the number of ways in which $n_j$ bosons can occupy two energy levels. The APF can then be written in terms of the original occupation vector $\vec{n}\vert_{N}$ by inserting a frequency factor to account for the extra level $j$: 
\begin{align}
\label{eq:Zbnj}
    \ZZZ B N j & =\sum_{\vec{n}\vert_{N}}(n_j+1)X(\vec{n}\vert_{N}) \\
               &= \Z{B}{N}\avg B N {n_j+1} \nonumber
\end{align}
and thus we can write:
\begin{equation}
\avg BN  {n_j} =\ZZZ B {N} j /\Z B N-1
\label{eq:njp1B}.
\end{equation}
Applying Eq.~(\ref{eq:ZNB_inv}), gives $\ZZZ F {N} j=\Z B N+\e{\epsilon_j}\ZZZ F {N-1} j$ which can be substituted into Eq.~(\ref{eq:njp1B}) to arrive at
\begin{equation}
    \avg BN  {n_j} =\frac{\e{\epsilon_j}\ZZZ B {N-1} j}{\Z B N}
\label{eq:njB},
\end{equation}
which is in the same form as Eq.~(\ref{eq:njF}) for fermions. 

To generalize this expression to $\ell$-level correlations with $\ell>1$ we examine the numerator of Eq.~(\ref{eq:njB}), recalling that we have added an extra copy of energy level $j$ to the partition function for $N-1$ particles such that it now appears $m_j+1=2$ times in the associated spectrum:
\begin{align}
    \e{\epsilon_j}
\ZZZ{B}{N-1}{j}& =\e{\epsilon_j}\sum_{\vec{n}^{\prime}\vert_{N-1}}
X(\vec{n}^{\prime}\vert_{N-1}) \nonumber \\
               &= \!\!\sum_{\vec{n}\vert_{N-1}}\!\!\binom{\tilde{n}_j + (m_j+1)-1}{(m_j+1)-1}    
                    \e{\epsilon_j}  X(\vec{n}\vert_{N-1}) \nonumber \\
               &= \sum_{\vec{n}\vert_{N, n_j\geq 1}}n_jX(\vec{n}\vert_{N})\,.
\label{eq:ZBANm1}
\end{align}
Here we obtain the second line using the same trick as in Eq.~(\ref{eq:Zbnj}) to convert $X(\vec{n}^\prime\vert_{N-1})$ into $X(\vec{n}\vert_{N-1})$ by accounting for the degeneracy where $[\vec{n}\vert_{N-1}]_j = \tilde{n}_j\geq 0$ is the total number of particles occupying the level $j$ as it appears in the Boltzmann factor $X(\vec{n}\vert_{N-1})=\e{\epsilon_j \tilde{n}_j}\prod_{i\neq j}\e{\epsilon_in_i}$.  The complicated looking binomial coefficient  
$\binom{\tilde{n}_j + (m_j+1)-1}{(m_j+1)-1} = \multiset{\tilde{n}_j}{m_j+1}$ is the multiset coefficient that counts the number of ways $\tilde{n}_j$ bosons can be distributed amongst the $m_j+1$ levels with energy $\epsilon_j$. Finally, the last line is obtained by using the 
fact that $\e{\epsilon_j}X(\vec{n}\vert_{N-1}) = X(\vec{n}\vert_N)$ where $[\vec{n}\vert_N]_j = n_j=\tilde{n}_j+1\geq 1$.  

Eq.~(\ref{eq:ZBANm1}) can be immediately extended to the case where we add copies of not 1 but $\ell$ levels $\qty{j_1,\dots,j_\ell}$:
\begin{equation}
% \e{(\epsilon_{j_1}+ \dots + \epsilon_{j_\ell})}
\e{\sum_{r=1}^\ell \epsilon_{j_r}}
\ZZZ B {N-\ell}{j_1, \dots, j_\ell} 
= \sum_{\vec{n}\vert_{N}}\qty(\prod_{r=1}^\ell {n}_{j_r})
% \tilde{n}_{j_2}\dots \tilde{n}_{j_\ell}\right)
% \e{\sum_{r=1}^\ell \epsilon_{j_r}}
X(\vec{n}\vert_{N})
% \e{(\epsilon_{j_1}+ \dots +\epsilon_{j_\ell})}
% \sum_{\vec{n}\vert_{N-\ell}}\left(\tilde{n}_{j_1}\tilde{n}_{j_2}\dots \tilde{n}_{j_\ell}\right)X(\vec{n}\vert_{N-\ell})
\label{eq:ZBANm1Gen}
\end{equation}
where the conditions $n_{j_r} \ge 1$ in the occupancy vector can be neglected as any $n_{j_r} = 0$ terms do not contribute to the sum due to the multiplicative string.  Finally, we can write the desired result:
\begin{equation}
\avg B N {n_{j_1}n_{j_2}\dots n_{j_\ell}}=\frac{\e{\sum_{r=1}^\ell \epsilon_{j_r}}}{\Z B N}\ZZZ B {N-\ell}{j_1, j_2, \dots, j_\ell}
\label{eq:corrB}.
\end{equation}
An immediate extension of Eq.~(\ref{eq:corrB}) will turn out to be useful, which introduces an APF with higher-order degeneracy. We consider $m_{j_r}$ extra copies of the $r^{\rm th}$ level $\epsilon_{j_r}$ with $r\in\{1,\dots\ell\}$ and find:
\begin{multline}
    \avg{B}{N}{\prod_{r=1}^{\ell}{\binom{n_{j_r}-q_{j_r}+m_{j_r}}{m_{j_r}}}} = \\
    \frac{1}{\Z{B}{N}} \e{\sum_{r=1}^{\ell}\epsilon_{j_r}q_{j_r}}\ZZZ B {N-\sum_{r=1}^{\ell}q_{j_r}} 
{\jdd{1}{1},\dots, \jdd{1}{m_{j_1}}, \dots \jdd{\ell}{1},\dots, \jdd{\ell}{m_{j_\ell}}}
\label{eq:njpmmkBl}
\end{multline}
where the $q_{j_r} \le m_{j_r}$ allow for the added freedom of choice of how many particles are associated with each of the degenerate levels and allow us to write the left-hand side in terms of the desired occupations $n_j = [\vec{n}\vert_{N}]_j$. Note: to simplify notation we only include a superscript on the levels in the bosonic APF and only if we are adding more than one extra copy per original level.

Eqs.~(\ref{eq:corrFgen}) and (\ref{eq:njpmmkBl})  are the major results of this section, and demonstrate that for both fermions and bosons, $\ell$-level correlations can be written in terms of APFs for $N-\ell$ particles with $\ell$ energy levels removed (added) for fermions (bosons).

% ---------------------------------------------------------------------------------
\subsection{Recovering Known Results Via the APF Theory }

In this section,  we illustrate the utility of our auxiliary expressions in simplifying the derivation of known recursion relations that govern the canonical partition functions and  occupation numbers for fermions and bosons. 

Beginning with fermionic statistics, if we use the definition of $\avg F N {n_j}$ in Eq.~(\ref{eq:njF}) to substitute for $\ZZ{F}{N}{j} $ and $\ZZ{F}{N-1}{j}$ in Eq.~(\ref{eq:ZNF}), 
we obtain the well-known recursion relation for occupation numbers \cite{Schmidt:1989,Borrmann_etal:1999}
\begin{equation}
\avg F {N+1} {n_j}=\frac{\Z F N}{\Z F {N+1}}\e {\epsilon_j}\left(1-{\avg F N {n_j}}\right)
\label{eq:njFrecursive1}\,.
\end{equation}
$\avg{F}{N}{n_j}$ can be written explicitly in term of the partition functions by substituting for $\ZZ{F}{N-1}{j}$ in Eq.~(\ref{eq:njF}) using Eq.~(\ref{eq:ZNF_inv}) as:
\begin{equation}
\avg F N {n_j}=\frac{1}{\Z F N}\sum_{k=1}^{N}(-1)^{k-1}\e {\epsilon_j k}\Z F {N-k}
\label{eq:njF_ZN},
\end{equation}
and using the canonical condition $\sum_i\avg F N {n_i}=N$ we find the original 1993 result of Borrmann and Franke~\cite{BorrmannFranke:1993}:
\begin{equation}
\Z F N=\frac{1}{N} \sum_{k=1}^{N}(-1)^{k-1} C_k \Z F {N-k}
\label{eq:ZNrecursiveF},
\end{equation}
where $C_k=\sum_j \e {\epsilon_j k}$. 

Bosonic statistics can be treated in analogy to the fermionic case. Assigning the roles played by Eqs.~(\ref{eq:njF}), (\ref{eq:ZNF}) and (\ref{eq:ZNF_inv}) to Eqs.~(\ref{eq:njB}), (\ref{eq:ZNB_inv}) and (\ref{eq:ZNB})\footnote{Here we replace $\Z B{x}$ with $\ZZZ{B}{x}{j}$ and $\ZZ{B}{x}{j}$ with $\Z{B}{x}$ in (\ref{eq:ZNB_inv}) and (\ref{eq:ZNB})}, respectively, we obtain the known bosonic equivalents \cite{Schmidt:1989,Borrmann_etal:1999,BorrmannFranke:1993}: 
\begin{align}
\label{eq:njBrecursive1}\avg B {N+1} {n_j}&=\frac{\Z B N}{\Z B {N+1}}\e{\epsilon_j}\qty(1+\avg B N {n_j})\\
\label{eq:njB_ZN}\avg B N {n_j}&=\frac{1}{\Z B N}\sum_{k=1}^{N}\e{\epsilon_jk}\Z B {N-k}\\
\label{eq:ZNrecursiveB}\Z B N&=\frac{1}{N} \sum_{k=1}^{N} C_k\Z B{N-k}\,.
\end{align}

Due to the Pauli exclusion principle for fermionic statistics, the occupation number $\avg {F}{N}{n_j}$ of an energy level $j$ gives us direct access to the occupation probability distribution $\p{F}{n_j}$ of the level. Despite the absence of any such simplification in the bosonic case, the occupation probability distribution $\p B{n_j}$ of a single energy level can be related to the corresponding partition functions as
\begin{equation}
\p B{n_j}=\e{\epsilon_j n_j}\frac{\Z B {N-n_j}}{\Z B N}-\e{\epsilon_j (n_j+1)}\frac{\Z B {N-n_j-1}}{\Z B N}
\label{eq:pnjBZ},
\end{equation}
which is obtained by using \Eqref{eq:pnB} to substitute for $\ZZ{B}{N-n_j}{j}$ in \Eqref{eq:ZNB_inv}, after replacing $N$ with $N-n_j$ \cite{WeissWilkens:1997}. In summary, within the unified framework of APFs, it is straightforward to obtain most of the well-known general relations in the fermionic and bosonic canonical ensemble that were previously derived using a host of different methods. This highlights the utility of this approach as a unifying framework when studying $N$ indistinguishable non-interacting particles.

% ---------------------------------------------------------------------------------
\subsection{General Expressions for Probabilities and Correlations}

De facto, the APFs can also be used to generalize previous results, and in a form that is highly symmetric with respect to particle statistics. To accentuate this, let us introduce the notation:
\begin{equation}
    \zeta = 
    \begin{cases}
        +1 & \Leftrightarrow  B \Leftrightarrow \text{bosons} \\
        -1 & \Leftrightarrow  F \Leftrightarrow \text{fermions} 
    \end{cases}\,.
\label{eq:zetaNotation}
\end{equation}
Then, the recursive relations for energy level correlations can be obtained using Eqs.~(\ref{eq:ZNF}) and (\ref{eq:ZNB_inv}) for fermions and bosons, respectively. The number of initial values of correlations needed is equal to the number of the involved points, e.g., for the two-level correlations we find:

\begin{align}
    \avg{\zeta}{N+2} {n_in_j}&=\frac{\Z{\zeta}{N}}{\Z{\zeta}{N+2}}\e {(\epsilon_i+\epsilon_j)}\qty(1-\avg{\zeta}{N}{n_in_j}) \nonumber \\
    &\;  + \zeta\frac{\Z{\zeta}{N+1}}{\Z{\zeta}{N+2}}\qty(\e {\epsilon_i}+\e {\epsilon_j})\avg{\zeta}{N+1} {n_in_j}
\label{eq:ninjFrecursivezeta}
\end{align}
Further, for the set of levels $\mathcal{S}_\ell=\{j_1, \dots, j_\ell\}$, Eqs.~(\ref{eq:pn1tolF}) and (\ref{eq:pn1tolB}), define the joint probability distributions of the occupation numbers in terms of the APFs $\ZZs F{k} {\mathcal{S}_\ell}$  and  $\ZZs B{k} {\mathcal{S}_\ell}$, which can be re-expressed using \Eqref{eq:Z1Z1F_inv} and \Eqref{eq:Z1Z1B_inv} as
\begin{equation}
    \p{\zeta}{n_{j_1},  \dots, n_{j_\ell}} =\frac{\e{E_{\rm{tot}}}}{\Z{\zeta}{N}}\sum_{k=0}^{N-n_{\rm{tot}}}(-1)^k\Z{-\zeta}{k}(\mathcal{S}_\ell) \Z{\zeta}{N-n_{\rm{tot}}-k} \\
\label{eq:PZ1Z1zeta_inv}
\end{equation}
where, $E_{\rm{tot}}=\sum_{r=1}^\ell \epsilon_{j_r}n_{j_r}$ and $n_{\rm{tot}}=\sum_{r=1}^\ell  n_{j_r}$. To avoid confusion we note again the convention in use that $\Z{-\zeta}{k}(\mathcal{S_\ell}) = 0$ whenever $k$ exceeds the maximum number of particles set by $\mathcal{S_\ell}$ for fixed particle statistics.

Now, as the level occupation correlations of fermions and bosons are represented by the APFs $\ZZs{F}{k} {\mathcal{S}_\ell}$ and $\ZZZs B{k} {\mathcal{S}_\ell}$ in Eqs. (\ref{eq:corrF}) and (\ref{eq:corrB}),  we can write
\begin{equation}
    \avg{\zeta}{N}{n_{j_1}\dots n_{j_\ell}} =\frac{\e{\sum_{r=1}^\ell \epsilon_{j_r}}}{\Z{\zeta}{N}}\sum_{k=0}^{N-\ell}\zeta^k\Z{B}{k}(\mathcal{S}_\ell) \Z{\zeta}{N-\ell-k}
\label{eq:corrzetaZ}
\end{equation}
where we note that the $k$-particle bosonic partition function for the levels $\mathcal{S}_\ell$ appears in both fermionic and bosonic correlations.

Alternatively, the APFs $\ZZs F{k} {\mathcal{S}_\ell}$,  $\ZZs B{k} {\mathcal{S}_\ell}$ and $\ZZZs B{k} {\mathcal{S}_\ell}$ can be computed  recursively using  \Eqref{eq:ZNrecursiveF} and \Eqref{eq:ZNrecursiveB}.  This has the potential to simplify the calculation of the related full joint probability distribution, as it only requires calculating the corresponding APF with a number of particles in the range  $k=0,\dots,N$.

% .................................................
\subsubsection{Simplification for degenerate levels}
% .................................................

The expressions derived in the previous section have a very simple form when the involved levels are degenerate, i.e., if we consider correlations between the set of levels $\{j^{(0)},\dots, j^{(\ell-1)}\}$, where $\epsilon_{j^{(s)}}=\epsilon_{j}$  for $s\in \{0,\dots, \ell-1\}$.  The canonical partition function  of $N$ bosons in $\ell$ degenerate energy levels is 
\begin{equation}
    \Z{B}{N}(\{j^{(0)}, \dots, j^{(\ell-1)}\})=\tbinom{N+\ell-1}{\ell-1}\e{\epsilon_{j}N}\, ,
\label{eq:degenZBN}
\end{equation}
and thus such correlations for fermions and bosons can be computed directly from \Eqref{eq:corrzetaZ} as
\begin{equation}
    \avg{\zeta}{N}{n_{j^{(0)}}\dots n_{j^{(\ell-1)}}}=\frac{1}{\Z{\zeta}{N}}\sum_{k=\ell}^{N}\zeta^{k-\ell}\tbinom{k-1}{\ell-1}\e{\epsilon_{j}k} \Z{\zeta}{N-k}
\label{eq:corrzetaZd}
\end{equation}
where we have shifted the summation. The numerical complexity of calculating  \Eqref{eq:corrzetaZd} differs from that of \Eqref{eq:njF_ZN}  or \Eqref{eq:njB_ZN} by the number of multiplications and additions needed to calculate the extra factor $\tbinom{k-1}{\ell-1}$, which can be viewed as a polynomial in $k$ of degree $\ell-1$. Thus calculating \Eqref{eq:corrzetaZd} requires additional $\approx(N-\ell)(\ell-1)$ multiplications, and a similar number of  additions which still scales linearly with $N$ for a moderate value of $\ell$. 

%
% .................................................
\subsubsection{Expectation values of higher moments of degenerate levels}
% .................................................

Let us now focus on the case of bosons and revisit \Eqref{eq:njpmmkBl} considering a single energy level $\epsilon_j$:
\begin{equation}
\avg{B}{N}{{\tbinom{n_{j}-q+m_j}{m_j}}} \!\!=\! 
\frac{1}{\Z{B}{N}} \en q {\epsilon_{j}}\ZZZ B {N-q}
{\jdd{}{1},\dots, \jdd{}{m_{j}}}.
\label{eq:njpmmkBl_1}
\end{equation}
Similarly, using \Eqref{eq:Z1Z2}, we can write the APF $\ZZZ B {N-q} {\jdd{}{1},\dots, \jdd{}{m_{j}}}$ in terms of the system partition function and the bosonic APF $\Z B {N-q} (\{\jdd{ }{1},\dots, \jdd{ }{m_{j}}\})$ of $m_j$ degenerate levels. As a result, we obtain
%
% \begin{widetext}
%
\begin{equation}
\avg{B}{N}{{\tbinom{n_{j}-q+m_j}{m_j}}} = \frac{1}{\Z B N}\sum_{k=q}^{N}\tbinom{k+m_j-q-1}{m_j-1}\e {\epsilon_{j}k}\Z B {N-k}
\label{eq:njpmmkBl_2},
\end{equation}
%
% \end{widetext}
%
where $q\leq m_j$. Note that if we set both of $m_j=\ell$ and $q=\ell$, then comparing with \Eqref{eq:corrzetaZd} we see that
\begin{equation}
\avg{B}{N}{{\binom{n_{j^{(s)}}}{\ell}}} =\avg B N {n_{j^{(0)}}\dots n_{j^{(\ell-1)}}}
\label{eq:nj_choose_l},
\end{equation}
for $s\in \{0,\dots, \ell-1\}$. This demonstrates that the correlations between degenerate levels can be expressed in terms of moments of the occupation number of \emph{any} of the degenerate levels.  This also helps to simplify the calculation of such moments, for example, we can write 
\begin{equation*}
\avg{B}{N}{n_j^2}=\avg{B}{N}{{\tbinom{n_{j}}{2}}}+\avg{B}{N}{{\tbinom{n_{j}+1}{2}}}   
\label{eq:B2ndmoment}
\end{equation*}
which can be simplified using \Eqref{eq:njpmmkBl_2} as
\begin{equation}
\avg{B}{N}{n_j^2} = \frac{1}{\Z B N}\sum_{k=1}^{N}\qty(2k-1)\e {\epsilon_{j}k}\Z B {N-k}
\label{eq:nj2B}.
\end{equation}
In the same fashion, we can write  
\begin{equation*}
\avg{B}{N}{n_j^3}=\avg{B}{N}{{\tbinom{n_{j}}{3}}}+4\avg{B}{N}{{\tbinom{n_{j}+1}{3}}}+\avg{B}{N}{{\tbinom{n_{j}+2}{3}}}
\end{equation*}
and thus
\begin{equation}
\avg{B}{N}{n_j^3} = \frac{1}{\Z B N}\sum_{k=1}^{N}\qty(3k^2-3k+1)\e {\epsilon_{j}k}\Z B {N-k}
\label{eq:nj3B}.
\end{equation}
%

% ---------------------------------------------------------------------------------
\section{Decomposition of level correlations into occupation numbers}
\label{sec:III}
In the previous section, we illustrated that the joint probability distributions of the occupation numbers and corresponding level correlations can be represented by auxiliary partition functions. The APF is distinguished from the actual partition function of the $N$-particle system through either the inclusion or exclusion of a set of levels from/to the complete spectrum under study. The resulting complexity of performing an actual calculation thus depends on the size of the modified set as demonstrated by, e.g. Eq.~(\ref{eq:corrzetaZ}).

It is known that the resulting complexity can be reduced by relating higher-order correlations between non-degenerate levels to the related level-occupation numbers \cite{Schonhammer:2017, GiraudGrabschTexier:2018}, representing an approach similar to Wicks theorem which only holds in the grand canonical ensemble. In this section, we directly obtain many known results using the APF method and, more importantly,  generalize them to deal with degenerate energy levels. 

Before we introduce the general and systematic approach to this problem, let us return to Eqs.~(\ref{eq:corrFgen}) and (\ref{eq:njpmmkBl}) and consider two specific examples of increasing difficulty.

% ---------------------------------------------------------------------------------
\subsection{Examples}
\subsubsection{Two-level correlations}
Consider the expectation value of two bosonic energy levels $j_1$ and $j_2$ where 
$\epsilon_{j_1}\neq \epsilon_{j_2}$: $\avg B N  {n_{j_1}n_{j_2}}$. Employing \Eqref{eq:njpmmkBl} with $m_{j_1}=m_{j_2}=1$, $q_{j_1}=0$ and $q_{j_2}=1$, we find
\begin{equation}
\avg{B}{N}{\qty(n_{j_1}+1)n_{j_2}} = \frac{\e{\epsilon_{j_2}}}{\Z{B}{N}} \ZZZ B {N-1}{j_1,j_2},
\label{eq:Ex1B1}
\end{equation}
and upon exchanging the values of $q_{j_1}$ and $q_{j_2}$, we have
\begin{equation}
\avg{B}{N}{n_{j_1}\qty(n_{j_2}+1)} = \frac{\e{\epsilon_{j_1}}}{\Z{B}{N}} \ZZZ B {N-1}{j_1,j_2}
\label{eq:Ex1B2}.
\end{equation}
Next, eliminating the APF from the two equations yields our final result: 
\begin{equation}
\avg{B}{N}{n_{j_1}n_{j_2}} = -\frac{\ep{\epsilon_{j_1}}\avg{B}{N}{n_{j_1}} -\ep{\epsilon_{j_2}}\avg{B}{N}{n_{j_2}} }{\ep{\epsilon_{j_1}}-\ep{\epsilon_{j_2}}}
\label{eq:Ex1B3}.
\end{equation}

Similarly, if the levels are fermionic, we use \Eqref{eq:corrFgen} with $\qty(\gamma_{j_1}, \gamma_{j_2})=\qty(0, 1)$ and $\qty(1, 0)$ to find:
\begin{equation}
\avg{F}{N}{n_{j_1}n_{j_2}} = \frac{\ep{\epsilon_{j_1}}\avg{F}{N}{n_{j_1}} -\ep{\epsilon_{j_2}}\avg{F}{N}{n_{j_2}} }{\ep{\epsilon_{j_1}}-\ep{\epsilon_{j_2}}}
\label{eq:Ex1F1}.
\end{equation}

These known results \cite{Schonhammer:2017, GiraudGrabschTexier:2018} are thus obtainable within the APF approach with a few lines of algebra by generating a set of independent equations.  We now extend this idea to three energy levels.

% ---------------------------------------------------------------------------------
\subsubsection{Three-level correlations}
\label{sssec:threelevelcorr}

The previous example for bosons is modified by adding a third level $j_3$ with $m_{j_3}=1$ and $\epsilon_{j_ 3}$ that is different than both of $\epsilon_{j_1}$ and $\epsilon_{j_2}$. Setting $\qty(q_{j_1}, q_{j_2}, q_{j_3})= \qty(1, 0, 1)$ and $\qty(1, 1, 0)$ in \Eqref{eq:njpmmkBl} gives the two equations

\begin{align}
\label{eq:Ex2B1}
\avg{B}{N}{n_{j_1}\qty(n_{j_2}+1)n_{j_3}} &= \frac{\e{\qty(\epsilon_{j_1}+\epsilon_{j_3})}}{\Z{B}{N}} \ZZZ B {N-2}{j_1, j_2, j_3} \\
\avg{B}{N}{n_{j_1}n_{j_2}\qty(n_{j_3}+1)} &= \frac{\e{\qty(\epsilon_{j_1}+\epsilon_{j_2})}}{\Z{B}{N}} \ZZZ B {N-2}{j_1, j_2, j_3}
\label{eq:Ex2B2}
\end{align}
respectively. Solving for $\avg{B}{N}{n_{j_1}n_{j_2}n_{j_3}}$ leads to
\begin{equation}
\avg{B}{N}{n_{j_1}n_{j_2}n_{j_3}} = -\frac{\ep{\epsilon_{j_2}}\avg{B}{N}{n_{j_1}n_{j_2}} -\ep{\epsilon_{j_3}}\avg{B}{N}{n_{j_1}n_{j_3}} }{\ep{\epsilon_{j_2}}-\ep{\epsilon_{j_3}}}
\label{eq:Ex2B3}.
\end{equation}
which can be further broken down into single-level occupation numbers by application of Eq.~(\ref{eq:Ex1B3}).

A slightly modified approach can be used if two of the energy levels are degenerate,   
$\epsilon_{j_2}=\epsilon_{j_3} \ne \epsilon_{j_1}$.  As above, we denote degenerate levels via superscript and we relabel $j_2 = j_2^{(0)}$ and $j_3 = j_2^{(1)}$.
Then it is clear that $\avg{B}{N}{n_{j_1^{\phantom{()}}}\!\!n_{j_2^{(0)}}} =\avg{B}{N}{n_{j_1^{\phantom{()}}}\!\!n_{j_2^{(1)}}}$ and Eq.~(\ref{eq:Ex2B3}) is not immediately applicable.  However, this can be resolved by replacing one of the choices of $\qty(q_{j_1^{\phantom{(1)}}}\!\!\!\!, q_{j_2^{(0)}}, q_{j_2^{(1)}})$, say $\qty(1, 1, 0)$, with $\qty(0, 1, 1)$, which gives%
 \begin{multline}
     \avg{B}{N}{\!n_{j_1^{\phantom{()}}}\!\!\!n_{j_2^{(0)}}n_{j_2^{(1)}}\!} = 
\\
-\frac{\ep{\epsilon_{j_1}}\!
 \avg{B}{N}{\!n_{j_1^{\phantom{()}}}\!\!n_{j_2^{(1)}}\!}\!
     % \avg{B}{N}{n_{j_1}n_{j_3}} 
 -\ep{\epsilon_{j_2}}
 \avg{B}{N}{n_{j_2^{(0)}}n_{j_2^{(1)}}\!\!}}
     % \avg{B}{N}{n_{j_1}n_{j_3}} 
 % -\ep{\epsilon_{j_2}}\avg{B}{N}{n_{j_2}n_{j_3}} }
 {\ep{\epsilon_{j_1}}-\ep{\epsilon_{j_2}}}
 \label{eq:Ex2B4}.
 \end{multline}

We now turn to the general decomposition of $\ell$-level correlations into functions of the occupation numbers of the energy levels for both the non-degenerate and degenerate spectra.

% ---------------------------------------------------------------------------------
\subsection{A Systematic Approach}
In the following we show that Eqs.~\eqref{eq:Z1Z2}, \eqref{eq:Z1Z1F_inv} and \eqref{eq:Z1Z1B_inv} can be directly employed to systematically relate higher-order correlations to levels occupation numbers. We begin by considering the set of levels $\mathcal{S}_\ell=\{j_1, j_2, \dots, j_\ell\}$ and relate the corresponding $\ell$-point correlations to the $r$-point correlations for any nonempty subset of levels $\mathcal{S}_r=\{i_1, i_2, \dots, i_r\}\subset\mathcal{S}_\ell$.

Imposing fermionic level statistics on the spectrum $\mathcal{S}$,  we can relate the correlations $\avg F N {n_{j_1}n_{j_2}\dots n_{j_\ell}}$ to $\avg F N {n_{i_1}n_{i_2}\dots n_{i_r}}$ by relating their corresponding APFs, i.e.,  $\ZZs F {N-\ell}{\mathcal{S}_\ell}$ and $\ZZs F {N-r}{\mathcal{S}_r}$. This can be achieved by building $\ZZs F {N-1} {\mathcal{S}_r}$ via Eq.~(\ref{eq:Z1Z2}) to combine the APFs of $\mathcal{S}\setminus\mathcal{S}_\ell$ with that of $\mathcal{S}_\ell\setminus\mathcal{S}_r$ through
\begin{equation}
\ZZs F {N-r} {\mathcal{S}_r} = \sum_{k=0}^{\ell-r}\ZZs F k {\mathcal{S}_r} ( \mathcal{S}_\ell) \ZZs F {N-r-k} {\mathcal{S}_\ell}
\label{eq:corrF1}.
\end{equation}
We note the upper limit in the previous summation is $\ell-r$, where in general it should be $k_{\max}=\min(\ell-r, N-r)$ (\Eqref{eq:Z1Z2}).  This is because the fermionic APF $\ZZs  F k {\mathcal{S}_r} ( \mathcal{S}_\ell)$ cannot describe more than $\ell-r$ particles, as this is the number of levels that are in $\mathcal{S}_\ell\setminus\mathcal{S}_r$. Further, for $N<\ell$ the correlations $\langle n_{j_1}n_{j_2}\dots n_{j_\ell}\rangle_N = 0$ for general particle statistics. As a result, we only consider $\ell$-points correlations with $N\ge\ell$. Moreover, we can obtain  $\ZZs  F k {\mathcal{S}_r} ( \mathcal{S}_\ell)$  by removing the contribution of the levels $\mathcal{S}_r$ from the APF of  $\mathcal{S}_\ell$, using Eq.~(\ref{eq:Z1Z1F_inv}), as
\begin{equation}
\ZZs F k {\mathcal{S}_r} (\mathcal{S}_\ell)=\sum_{m=0}^{k}(-1)^{m}\Z B {m}(\mathcal{S}_r) \Z F {k-m}(\mathcal{S}_\ell)
\label{eq:corrF2}.
\end{equation}
If we isolate the last term in the summation in \Eqref{eq:corrF1}, i.e.,   $\ZZs F {\ell-r} {\mathcal{S}_r} ( \mathcal{S}_\ell) \ZZs F {N-\ell} {\mathcal{S}_\ell}$, while substituting for $\ZZs F k {\mathcal{S}_r} (\mathcal{S}_\ell)$, using Eq.~(\ref{eq:corrF2}) in all other terms, we find
\begin{align}
\ZZs F {N-r} {\mathcal{S}_r} =&\sum_{k=0}^{\ell-r-1}\sum_{m=0}^{k}(-1)^{m}\Z B {m}(\mathcal{S}_r) \Z F {k-m}(\mathcal{S}_\ell) \ZZs F {N-r-k} {\mathcal{S}_\ell}\nonumber\\
&+\ZZs F {\ell-r} {\mathcal{S}_r} ( \mathcal{S}_\ell) \ZZs F {N-\ell} {\mathcal{S}_\ell}
\label{eq:corrF3}.
\end{align}
After changing the order of the summations and shifting the indexes  $k\to k-r+1$ and $m\to m-r$, we obtain the unwieldy expression 
\begin{align}
    \ZZs F {N-r} {\mathcal{S}_r} &=(-1)^{r-1}\!\!\sum_{m=r}^{\ell-1}(-1)^{m+1}\Z B {m-r}(\mathcal{S}_r)\!\!\! \! \nonumber \\
                              &\times \sum_{k=m-1}^{\ell-2}\!\!\!\!\Z F {k-m+1}(\mathcal{S}_\ell) \ZZs F {N-k-1} {\mathcal{S}_\ell}\nonumber\\
&+\ZZs F {\ell-r} {\mathcal{S}_r} ( \mathcal{S}_\ell) \ZZs F {N-\ell} {\mathcal{S}_\ell}
\label{eq:corrF4}.
\end{align}
Next, we substitute for $\ZZs F {N-\ell} { \mathcal{S}_\ell}$ and $\ZZs F {N-r} {\mathcal{S}_r}$, using Eq.~(\ref{eq:corrF}) and the fully occupied fermionic APF $\ZZs F {\ell-r} {\mathcal{S}_r} ( \mathcal{S}_\ell) =\e {\sum_{j_\nu\in\mathcal{S}_\ell\setminus\mathcal{S}_r}\epsilon_{j_\nu}}$. Multiplying the result by $\e {\sum_{i_\nu\in\mathcal{S}_r}\epsilon_{i_\nu}}/\Z{F} {N}$ yields:
\begin{equation}
\Y F 0(\mathcal{S}_\ell)+\sum_{m=r}^{\ell-1}A_{m}(\mathcal{S}_r)\Y F m(\mathcal{S}_\ell)=b_{F}(\mathcal{S}_r)
\label{eq:LinearF}
\end{equation}
where
\begin{align}
\Y F 0(\mathcal{S}_\ell)&=\avg F N {n_{j_1}n_{j_2}\dots n_{j_\ell}} \\
\Y F {1\le m\le\ell-1}(\mathcal{S}_\ell)&=
\frac{(-1)^{m+1}}{\Z F N}\!\!\!\! \sum_{k=m-1}^{\ell-2}\!\!\!\!\Z F {k-m+1}(\mathcal{S}_\ell) \ZZs F {N-k-1} {\mathcal{S}_\ell} 
\label{eq:YmF}
\end{align}
are independent of $\mathcal{S}_r$.  Therefore, for each choice of the subset
$\mathcal{S}_r$ we can write the linear nonhomogeneous equation
\eqref{eq:LinearF} in the $\ell$ variables $\Y F m$ with coefficients: 
$A_{0}=1$, $A_{ 0<m<r}=0$ and $A_{ r\le m \le\ell-1}(\mathcal{S}_r)=(-1)^{r-1}\e {\sum_{i_\nu\in\mathcal{S}_r}\epsilon_{i_\nu}}\Z B {m-r}(\mathcal{S}_r)$. The homogeneity of the linear equation is violated by the term $b_F(\mathcal{S}_r)=\avg F N {n_{i_1}n_{i_2}\dots n_{i_r}}$.

With this formulation, we observe that for any of the $2^\ell-2$ choices of $\mathcal{S}_r$, we can write a linear equation in the same $\ell$ variables $\Y F m$, where $\Y F 0=\avg F N {n_{j_1}n_{j_2}\dots n_{j_\ell}}$ is the $\ell$-point correlation while the remaining $\ell-1$ variables are auxiliary, and depend symmetrically on the levels in $\mathcal{S}_\ell$. Also, the $r$-point correlation $\avg F N {n_{i_1}n_{i_2}\dots n_{i_r}}$ of the levels in $\mathcal{S}_r$ plays the role of the nonhomogeneous term in the linear equation and the coefficients $A_{m}$ of the equation can be determined by the bosonic APFs of  $\mathcal{S}_r$. An analogous expression can be obtained for bosonic statistics, with the same coefficients $A_{m}$ 
\begin{equation}
\Y B 0(\mathcal{S}_\ell)+\sum_{m=r}^{\ell-1}A_{m}(\mathcal{S}_r)\Y B m(\mathcal{S}_\ell)=b_{B}(\mathcal{S}_r)
\label{eq:LinearB},
\end{equation}
where, in this case, the variables are 
\begin{multline}
\Y B{1 \le m \le \ell-1}(\mathcal{S}_\ell)= \\
\frac{(-1)^{m+1}}{\Z B N} \sum_{k=m-1}^{\ell-2}(-1)^{k}\Z F {k-m+1}(\mathcal{S}_\ell) \ZZZs B {N-k-1} {\mathcal{S}_\ell}
\label{eq:YmB},
\end{multline}
with $\Y B 0(\mathcal{S}_\ell)=\qty(-1)^{\ell-1}\avg B N {n_{j_1}n_{j_2}\dots n_{j_\ell}}$ and $b_B(\mathcal{S}_r)=(-1)^{r-1}\avg B N {n_{i_1}n_{i_2}\dots n_{i_r}}$ (see Appendix~\ref{Appendix:A} for a complete derivation).

% ...........................................................................
%Non-degenerate case
\subsubsection{Non-degenerate levels}
% ...........................................................................
Consider the set $\mathcal{S}_\ell$ specifying a set of distinct energy levels and choose $\mathcal{S}_{r=1}$ such that it contains only one of the $\ell$ levels in $\mathcal{S}_\ell$.  We can use Eq.~(\ref{eq:LinearF}) or (\ref{eq:LinearB}) to construct a set of $\ell$ linear equations each corresponding to one level $j_s\in\mathcal{S}_\ell$ with energy $\epsilon_{j_s}$. For fermions the equations are
\begin{equation}
\avg F N  {n_{j_s}} =\sum_{m=0}^{\ell-1}\en m{\epsilon_{j_s}}\Y F m(\mathcal{S}_\ell)
\label{eq:corrF1non-deg},
\end{equation}
where the coefficients $A_m$ were obtained from the single-level bosonic APF $\Z B {m}(\{j_s\})=\en m {\epsilon_{j_s}}$ in \Eqref{eq:ZjBosons}. Therefore, using the set of the $\ell$ independent linear equations in $\ell$ variables defined by Eq.~(\ref{eq:corrF1non-deg}), we can solve for $\Y F 0(\mathcal{S}_\ell)=\avg F N {n_{j_1}n_{j_2}\dots n_{j_\ell}}$ as
\begin{align}
\avg F N {n_{j_1}\dots n_{j_\ell}} &=
\frac{\begin{vmatrix}
\avg F N  {n_{j_1}} & \e{\epsilon_{j_1}} & \dots & \e{(\ell-1)\epsilon_{j_1}} \\ 
\vdots & \vdots & \ddots  & \vdots \\ 
\avg F N  {n_{j_{\ell}}} & \e{\epsilon_{j_{\ell}}} & \dots & \e{(\ell-1)\epsilon_{j_{\ell}}} \\  
\end{vmatrix}}
{\begin{vmatrix}
1 & \e{\epsilon_{j_1}} & \dots & \e{(\ell-1)\epsilon_{j_1}} \\ 
\vdots & \vdots & \ddots  & \vdots \\ 
1 & \e{\epsilon_{j_{\ell}}} & \dots & \e{(\ell-1)\epsilon_{j_{\ell}}}
\end{vmatrix}}
\label{eq:corrFSol1}.
\end{align}

This result was recently obtained by Giraud, Grabsch and Texier \cite{GiraudGrabschTexier:2018}, using the properties of the Schur functions and it can be simplified using Vandermonde determinants:
\begin{align}
    \avg{F}{N}{n_{j_1}n_{j_2}\dots n_{j_\ell}} &= \sum_{s=1}^\ell(-1)^{s-1}
    \avg{F}{N}{n_{j_s}}\e{\sum_{j_i\neq j_s}\epsilon_{j_i}} \nonumber \\
                                               &\quad \times \frac{\mathbb{V}^{\setminus\{j_s\}}\qty(\e{\epsilon_{j_1}},\dots,\e{\epsilon_{j_{\ell}}})}{\mathbb{V}\qty(\e{\epsilon_{j_1}},\dots,\e{\epsilon_{j_{\ell}}})}
\label{eq:corrFSol3},
\end{align}
where 
\begin{equation}
\mathbb{V}\qty(\alpha_1,\dots,\alpha_\ell) =
\begin{vmatrix}
1 & \alpha_1 & \dots & \alpha_1^{\ell-1} \\ 
\vdots & \vdots & \ddots  & \vdots \\ 
1 & \alpha_\ell & \dots & \alpha_\ell^{\ell-1}
\end{vmatrix} 
=\prod_{i<j}\qty(\alpha_j-\alpha_i)
\label{eq:Vandermonde}.
\end{equation}
% and can be simplified as $\mathbb{V}\qty(\alpha_1,\dots,\alpha_n)=\Pi_{i<j}\qty(\alpha_j-\alpha_i)$. 
Thus the fermionic $\ell$-level correlation can be simplified as: 
\begin{equation}
\avg F N {n_{j_1}n_{j_2}\dots n_{j_\ell}}=\sum_{r=1}^{\ell}\frac{\avg F N  {n_{j_r}}}{\prod_{k\neq r}\qty[1-\ep{\qty(\epsilon_{j_k}-\epsilon_{j_r})}]}
\label{eq:corrFSol}.
\end{equation}
This expression was also recently derived using an elegant second quantization scheme\cite{Schonhammer:2017}.

An equivalent procedure can be performed for bosons, again using the set $\mathcal{S}_{r=1}$ to yield $\ell$ linearly independent equations
\begin{equation}
\avg B N  {n_{j_r}} =\sum_{m=0}^{\ell-1}\en m{\epsilon_{j_r}}\Y B m(\mathcal{S}_\ell)
\label{eq:corrB1non-deg},
\end{equation}
such that $\avg B N {n_{j_1}n_{j_2}\dots n_{j_\ell}}=\qty(-1)^{\ell-1}\Y B 0(\mathcal{S}_\ell)$ can also be expressed in terms of determinants
leading to\cite{GiraudGrabschTexier:2018}
\begin{equation}
\avg B N {n_{j_1}n_{j_2}\dots n_{j_\ell}}=\qty(-1)^{\ell-1}\sum_{r=1}^{\ell}\frac{\avg B N  {n_{j_r}}}{\prod_{k\neq r}\qty[1-\ep{\qty(\epsilon_{j_k}-\epsilon_{j_r})}]}
\label{eq:corrBSol}.
\end{equation}
%

% ...........................................................................
%Degenerate case
\subsubsection{Degenerate levels}
% ...........................................................................

Up until this point we have considered $\ell$-point level correlations in two opposite regimes: (I) when all $\ell$-levels are degenerate Eq.~(\ref{eq:corrzetaZd}) provides a direct route to the correlations through the determination of all partition functions up to $N$ particles, and (II) when all $\ell$ levels are distinct, an associated set of $\ell$ linear equations yields the correlations in terms of individual level occupation numbers.  The independence of these linear equations, and thus the existence of a unique solution, is violated in the presence of degeneracy. 

We now study the most general possible $\ell$-level correlation function defined by the set $\mathcal{S}_\ell$ which could include both degenerate and non-degenerate levels.  Consider the subset of level indices $\mathcal{S}_{m_i}=\{i^{(0)}, \dots, i^{(m_i-1)}\} \subset \mathcal{S}_\ell$ which contains $m_i > 1$ degenerate levels   (for $m_i = 1$, we reproduce the non-degenerate analysis discussed above).  As a result, the corresponding $m_i$  equations, out of the total set of $\ell$ linear equations defined by Eqs.~(\ref{eq:corrF1non-deg}) and \eqref{eq:corrB1non-deg} for fermions and bosons respectively are identical, as they are distinguished from each other only via the energies of the involved levels and their occupation numbers.   Moreover, $\mathcal{S}_\ell$, could contain multiple subsets of degenerate energy levels, further complicating the problem.  Eqs.~\eqref{eq:corrFSol} and \eqref{eq:corrBSol} can not be applied in this case, as is apparent from their vanishing denominators whenever $\epsilon_{j_k}=\epsilon_{j_r}$. 

To resolve the complication introduced by degenerate subsets, we generalize the procedure in \S\ref{sssec:threelevelcorr} to treat the case where a three-level correlation contained a subset of 2 degenerate levels.  More explicitly, we relate the $\ell$-points correlations not only to the occupation numbers of the degenerate subsets, but to all of the $m_i$ distinct $r$-points correlations between the degenerate levels with $2 \le r\le m_i$. This is useful as we have already introduced \Eqref{eq:corrzetaZd}, which simplifies the calculation of the correlations between degenerate energy levels for fermionic and bosonic statistics. In addition, it will result in the generation of $m_i$ new independent equations that could be used to calculate $\ell$-points correlations. 

Accordingly, for any choice of $\mathcal{S}_{r}=\{i^{(0)}, \dots, i^{(r-1)}\}\subset \mathcal{S}_{m_i}\subset \mathcal{S}_{\ell}$, the corresponding coefficients in the constructed linear equations are $A_{0}=1$, $A_{0<m<r}=0$ and 
\begin{equation}
A_{r\le m \le \ell-1}(\mathcal{S}_r)= (-1)^{r-1}\binom{m-1}{r-1} \en  m {\epsilon_i}
\label{eq:AFdeg},
\end{equation}
where $\epsilon_i$ is the energy of all degenerate levels in $\mathcal{S}_{m_i}$ and 
we have substituted for $\Z B  {m-r}(\{i^{(0)}, \dots, i^{(r)}\})=\binom{m-1}{r-1}\en {(m-r)}{\epsilon_{i}}$, a bosonic partition function of $r$ degenerate levels. 

The non-homogeneities $b_\zeta(\mathcal{S}_r)=(-\zeta)^{r-1}\avg{\zeta}{N }{n_{i^{(1)}}n_{i^{(2)}}\dots n_{i^{(r-1)}}}$ can be   
calculated using \Eqref{eq:corrzetaZd}, and thus, the original set of $m_i$ identical equations, can now be replaced with the following $m_i$ independent equations
\begin{equation}
    \Y{\zeta}{0}(\mathcal{S}_\ell)+\sum_{m=r}^{\ell-1}(-1)^{r-1}\binom{m-1}{r-1}\en m{\epsilon_{i}}\Y{\zeta}m(\mathcal{S}_\ell)=b_\zeta(\mathcal{S}_r)
\label{eq:LinearFdeg},
\end{equation}
for fermions $(\zeta=-1)$ and bosons $(\zeta=+1)$.

To illustrate how this works in practice, consider the $4$-point correlation of the levels $\{j_{1^{(0)}}, j_{1^{(1)}}, j_{1^{(2)}}, j_2^{\phantom{()}}\!\!\}$, labeling distinct energies $\epsilon_1$ and $\epsilon_2$. The resulting new set of $\ell$ equations can be solved for both fermions and bosons to give:
\begin{widetext}
\begin{align}
    \avg{\zeta}{N}{n_{j_{1^{(0)}}}n_{j_{1^{(1)}}}n_{j_{1^{(2)}}}n_{j_2^{\phantom{()}}\!\!}} &= (-\zeta)
\frac{\begin{vmatrix}
\avg \zeta N {n_{j_{1^{(0)}}}n_{j_{1^{(1)}}}n_{j_{1^{(2)}}}} &0&0 & \en 3 {\epsilon_{j_{1}}} \\ 
(-\zeta)\avg \zeta N {n_{j_{1^{(0)}}}n_{j_{1^{(1)}}}} &0&-\en 2 {\epsilon_{j_{1}}} & -2\en 3 {\epsilon_{j_{1}}} \\ 
\avg \zeta N  {n_{j_{1^{(0)}}}} &\e{\epsilon_{j_{1}}}&\en 2 {\epsilon_{j_{1}}} & \en 3 {\epsilon_{j_{1}}} \\ 
\avg \zeta N  {n_{j_{2}^{\phantom{()}}}\!\!} &\e{\epsilon_{j_{2}}}&\en 2 {\epsilon_{j_{2}}} & \en 3 {\epsilon_{j_{2}}} \\ 
\end{vmatrix}}
{\begin{vmatrix}
1&0&0 & \en 3 {\epsilon_{j_{1}}} \\ 
1 &0&-\en 2 {\epsilon_{j_{1}}} & -2\en 3 {\epsilon_{j_{1}}} \\ 
1 &\e{\epsilon_{j_{1}}}&\en 2 {\epsilon_{j_{1}}} & \en 3 {\epsilon_{j_{1}}} \\ 
1&\e{\epsilon_{j_{2}}}&\en 2 {\epsilon_{j_{2}}} & \en 3 {\epsilon_{j_{2}}}  \\ 
\end{vmatrix}}
\label{eq:Exzetadeg}.
\end{align}
\end{widetext}
%%

% ---------------------------------------------------------------------------------
\section{Applications}
\label{sec:IV}
% ---------------------------------------------------------------------------------
%
To illustrate the applicability of our results for degenerate non-interacting systems of particles with fixed number, and highlight the practical usage of Eq.~(\ref{eq:Exzetadeg}) we consider a one-dimensional tight-binding chain of $N$ spinfull bosons hopping over $L$ lattice sites.  Code, scripts, and data used to produce all figures for the bosonic chain results can be found online \cite{repo}.  The bosonic chain is subject to a static external magnetic field $B$ applied along the $z$-axis. Spin-$S$ bosons are described by the Hamiltonian
\begin{equation}
    \widehat{H}=-t\sum_{\alpha,\sigma} \left( \hat{a}^\dag_{\alpha+1,\sigma}\hat{a}^{\phantom{\dag}}_{\alpha,\sigma}+\text{h.c.}\right) -h\sum_{\alpha,\sigma,\sigma^\prime }\hat{a}^\dag_{\alpha,\sigma}{S}^{z}_{\sigma,\sigma^\prime}\hat{a}^{\phantom{\dag}}_{\alpha,\sigma^\prime}
\label{eq:B_Ring},
\end{equation}
where $\hat{a}^\dag_{\alpha,\sigma}$ and  $\hat{a}^{\phantom{\dag}}_{\alpha,\sigma}$ are creation and annihilation operators for a boson at site $\alpha$ with $\sigma\in\{-S,\dots, 0, \dots, S\}$ satisfying $[a^{\phantom{\dag}}_{\alpha,\sigma},a^\dag_{\alpha^\prime,\sigma^\prime}] = \delta_{\alpha,\alpha^\prime}\delta_{\sigma,\sigma^\prime}$ and $t$ measures the hopping amplitude. ${S}^{z}_{\sigma,\sigma^\prime} = \sigma \delta_{\sigma,\sigma^\prime}$ are the matrix elements of the diagonal $z$-projection of the spin-$S$ representation of the spin operator $\hat{\vec{S}}$. Here, $h=g{\rm{\mu_B}}B$, where $g$ is the corresponding spin-$S$ $g$-factor and $\mu_B$ is the Bohr magneton.  We employ periodic boundary conditions, such that $\hat{a}^{\phantom{\dag}}_{L+1,\sigma}=\hat{a}^{\phantom{\dag}}_{1,\sigma}$, and to avoid having an unbalanced non-degenerate excited state, we fix the parity of $L$ to be odd. 

The tight-binding Hamiltonian in Eq.~(\ref{eq:B_Ring}) can be diagonalized 
\begin{equation}
    \widehat{H}=\sum_{j,\sigma}\epsilon_{j,\sigma} n_{j,\sigma}
\label{eq:B_Ring_1p_diagonal},
\end{equation}
where $\hat{n}_{j,\sigma}$ counts the number of bosons with energy  
% = \hat{\tilde{a}}^\dag_{j,\sigma}\hat{\tilde{a}}^{\phantom{\dag}}_{j,\sigma}$
%
\begin{equation}
\epsilon_{j,\sigma}=-2t\cos\qty(\frac{2\pi j}{L})-h\sigma
\label{eq:SPE},
\end{equation}
and $j$ runs over the finite set $\mathcal{S} = \{-\frac{L-1}{2}, \dots, 0, \dots, \frac{L-1}{2}\}$ when $L$ is odd such that $\abs{\mathcal{S}} = L$. An examination of the single-particle spectrum shows that each energy level, except the ground state $\epsilon_{0,\sigma}=-2t-h\sigma$, is $2$-fold degenerate, where $\epsilon_{-j,\sigma}=\epsilon_{j,\sigma}$; a result of the right-left symmetry of the chain. Turning off the magnetic field and fixing $S>0$, gives rise to an extra degeneracy factor of $(2S+1)$ that affects all levels. 

For all numerical results presented in this section, we fix $L=1001$, $N=1000$ and measure the inverse temperature $\beta=\frac{1}{k_{\rm B}T}$ in units of $1/t$, where $T$ is the absolute temperature and $k_{\rm{B}}$ is the Boltzmann constant.
% ---------------------------------------------------------------------------------
\subsection{Spinless bosons $(S=0)$}
We begin with the study of spinless bosons, where the model is insensitive to the applied magnetic field and we can drop the subscript $\sigma$ without loss of generality. Using the single-particle spectrum defined in \Eqref{eq:SPE} with $\sigma=0$ and $h=0$ 
in combination with \Eqref{eq:pn1tolB}, we calculate the joint probability distribution $\p{B}{n_{0},n_{1}}$ of the occupation numbers of the ground state and the first excited state, where, we choose the level $j=1$ out of the two degenerate levels $j=\pm1$. Note that we do not bother to use the superscript notation to distinguish degenerate level indices $(1\equiv 1^{(0)},-1\equiv1^{(1)})$ here as there are no ambiguities due to the sign of the index $j$. 

The calculation proceeds by obtaining the APFs $\ZZ{B}{k}{0,1}$ for $0\leq k\leq N$ using the recursion relation \Eqref{eq:ZNrecursiveB}, where the factors $C_k$ are calculated using the spectrum $\mathcal{S} \setminus \qty{0,1}$. The resulting distribution is 
\begin{equation}
    \p B{n_{0}, n_{1}}=\frac{\e{\qty(\epsilon_{0}n_{0}+\epsilon_{1}n_{1})}}{\Z B N} \ZZ B {N-n_{0}-n_{1} } {0,1}
\label{eq:pn01}.
\end{equation}
where $\Z B N$ can be found by enforcing normalization.

\begin{figure}[t]
\begin{center}
\includegraphics[width=1.0\columnwidth]{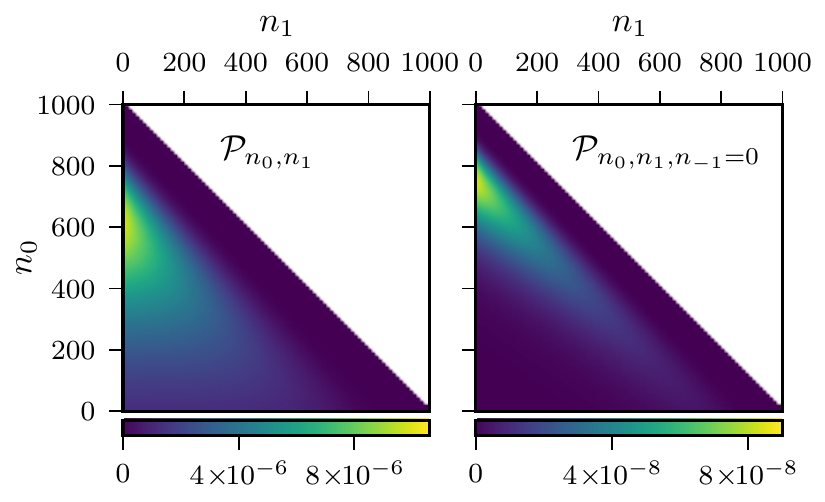}
\end{center}
\caption{The joint level probability distribution for $N=1000$ spinless bosons on chain of $L=1001$ sites described by Eq.~(\ref{eq:B_Ring_1p_diagonal}) with $\sigma=0$ and $h=0$ at inverse temperature $\beta=100/t$. Left panel: $\p{B}{n_{0},n_{1}}$.  Right panel:  $\p{B}{n_{0},n_{1},n_{-1}}$ projected into the plane $n_{-1}=0$.}
\label{fig:probbeta100}
 \end{figure}

We expect Eq.~(\ref{eq:pn01}) to exhibit interesting features at low temperature where the particles are mostly occupying the ground state with some fluctuations amongst the low lying energy levels.  To obtain an estimate of \emph{low} in this context, we choose a value of the inverse temperature $\beta$ such that the ground state has a macroscopic occupation corresponding to 50\% of the particles. We compare the ratio of the Boltzmann factors of having all particles in the ground state with that of having $N/2$ particles in the first excited state and the rest in the ground state. Setting the ratio of these factors $\e {\qty(\epsilon_0-\epsilon_1)N/2}$ to $\sim 0.1$, suggests $\beta\sim 100/t$. The results are illustrated in the left panel of Fig.~\ref{fig:probbeta100},
%%%%%%%%%% 
%%%%%%%%%%
where the relative broadness of the distribution can be attributed to the degeneracy of the first exited level $j=\pm 1$. 

If we now calculate the three-level joint probability distribution $\p{B}{n_{0},n_{1},n_{-1}}$ and consider the fixed slice with $n_{-1}=0$, as presented in the right panel of Fig.~\ref{fig:probbeta100}, we see that the distribution becomes significantly sharper, as blocking the level $j=-1$ makes the resulting non-normalized conditional distribution more sensitive to the conservation of the total number of particles.

We now turn to the calculation of the two-level connected correlation function for our bosonic system
\begin{equation}
\mathcal{C}(n_i,n_j)=\avg B N {n_in_j}-\avg B N {n_i}\avg B N {n_j}. 
\label{eq:Cninj}
\end{equation}
The first step is to obtain the system partition function $\Z B k$, recursively, using \Eqref{eq:ZNrecursiveB} starting from $\Z B 0$ up to $\Z B N$. The occupation numbers $\avg B N {n_j}$ can then be easily calculated using \Eqref{eq:njB_ZN}. All that remains is to calculate the two-points correlations using \Eqref{eq:Ex1B3} for the non-degenerate levels. For the correlations between degenerate levels ($\avg B N {n_{-j}n_j}$), we use \Eqref{eq:corrzetaZd}, with $\zeta=+1$. 

\begin{figure}[t]
\begin{center}
\includegraphics[width=1.0\columnwidth]{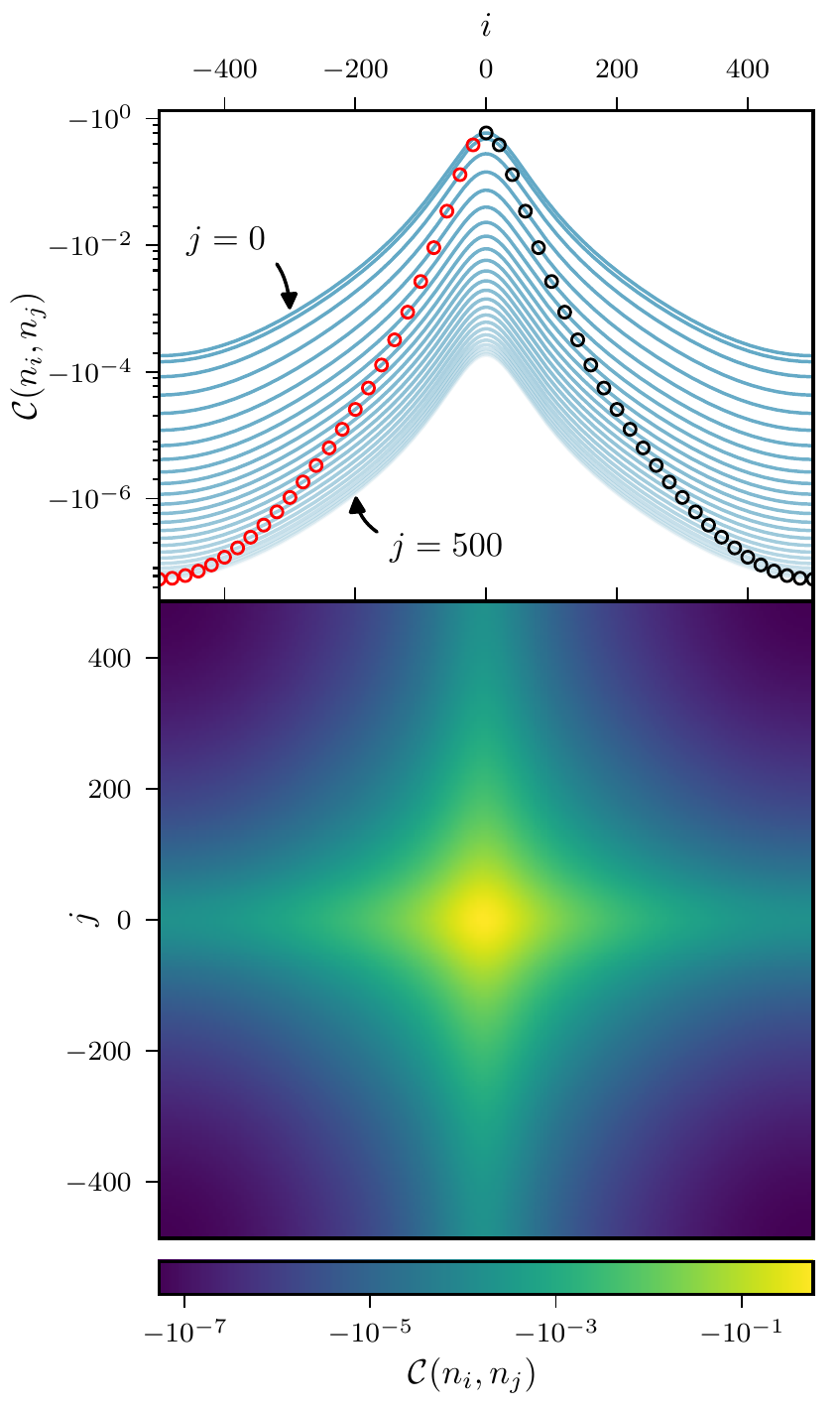}
\end{center}
\caption{Lower panel: Heat-map depicting two-points connected correlations $\mathcal{C}(n_i,n_j)=\avg B N {n_in_j}-\avg B N {n_i}\avg B N {n_j}$ in a system of $N=1000$ spinless bosons on a lattice of $L=1001$ sites at $\beta=1/t$. Upper panel: Horizontal cuts from the upper half of $\mathcal{C}(n_i,n_j)$ at different values of the index $j>0$. The circled data points are calculated using \Eqref{eq:corrzetaZd}, where the red-circled points are $Cr(n_{-j},n_j)$ for the degenerate levels $j$ and $-j$, while the black-circled points are $\lim_{i\to j}\mathcal{C}(n_{i},n_j)=\avg B N {\tbinom{n_j}{2}}-{\avgsq B N {n_{j}}}$.}
\label{fig:Seq0}
 \end{figure}

The results for $\mathcal{C}(n_i,n_j)$ for all levels $i$ and $j$ at inverse temperature $\beta=1/t$ are shown as a heat-map in the lower panel of Fig.~\ref{fig:Seq0}.
%%%%%%%%%% .Figure~\ref{fig:S=0}
%%%%%%%%%%
Here, we chose temperature $\beta=1/t$ in order to distribute the correlations amongst higher energy levels. The upper panel of Fig.~\ref{fig:Seq0} shows $\mathcal{C}(n_i,n_j)$ as a function of $n_i$ for fixed $0\leq j \leq 500$ corresponding to horizontal cuts through the lower panel.  The red open circles are correlations $\mathcal{C}(n_{-j},n_j)$ for the degenerate levels 
 obtained from \Eqref{eq:corrzetaZd} demonstrating consistency with the rest of the graph. In the positive quadrant of the correlation heat-map,  we use the values of $\avg B N {\tbinom{n_j}{2}}$ (marked with black circles)  instead of $\avg B N {n_j^2}$, were the former is also consistent with surrounding data, as expected from \Eqref{eq:nj_choose_l}, where $\avg B N {n_{-j}n_{j}}=\avg B N {\tbinom{n_j}{2}}$ and the symmetry $\mathcal{C}(n_{i},n_j)=\mathcal{C}(n_{-i},n_j)$ due to the degeneracy.

% ---------------------------------------------------------------------------------
\subsection{Spin-$1$ bosons $(S=1)$}
To illustrate the utility of auxiliary partitions functions in studying correlations in a highly degenerate spectrum, we consider the case of spin-$1$ bosons. In the absence of a magnetic field ($h=0$), each level picks up a degeneracy factor of $2S+1=3$, such that the ground state is three-fold degenerate and all of the excitation levels are six-fold degenerate.

\begin{figure}[t]
\begin{center}
\includegraphics[width=1.0\columnwidth]{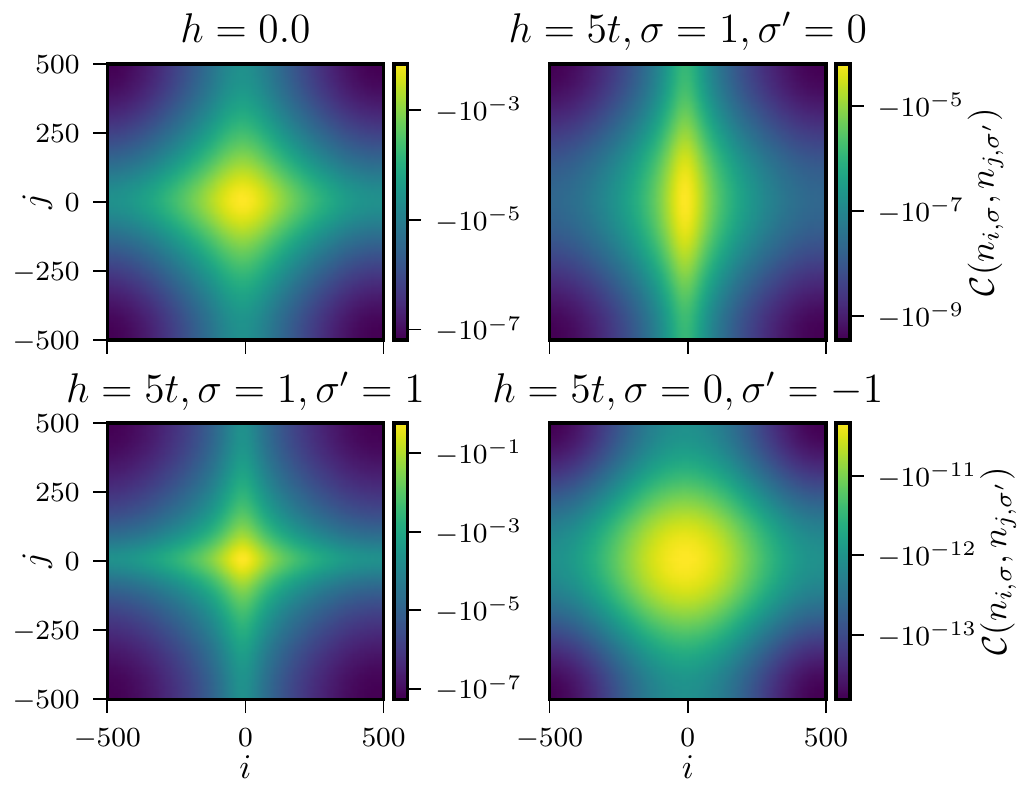}
\end{center}
\caption{Two-level connected correlation function $\mathcal{C}(n_{i,\sigma},n_{j,\sigma^\prime})$ at $\beta=1/t$ for $N=1000$ spin-1 bosons on $L=1001$ sites. Panels correspond to different values of $\sigma$, $\sigma^\prime$ and the applied magnetic field ($h=g{\rm{\mu_B}}B$) as indicated.}
\label{fig:Seq1}
 \end{figure}

Degeneracy effects are apparent in the two-level connected correlations $\mathcal{C}(n_{i,\sigma},n_{j,\sigma^\prime})$ at $\beta=1/t$, which we calculate for various values of $h$ as shown in Fig.\ref{fig:Seq1}.
%%%%%%%%%%
%%%%%%%%%%
%
The top-left panel of the figure presents $\mathcal{C}(n_{i,\sigma},n_{j,\sigma^\prime})$ for $h=0$, where the choice of $\sigma$ and $\sigma^\prime$ matters only in the presence of a magnetic field. A comparison with the heat-map of Fig.~\ref{fig:Seq0}, shows an overall broadening and a reduction of one order of magnitude in the maximum of $\mathcal{C}(n_{i,\sigma},n_{j,\sigma^\prime})$ for $S=1$, as compared to $S=0$ case. 

Removing the energy-spin degeneracy by applying a strong magnetic field of $h=5t$, results in a splitting of the spectrum into three bands, each with bandwidth $4t$ and separated from each other via an energy bandgap of $t$.  In this case, we first focus on correlations between the levels in the lower energy band ($\sigma=1$, bottom-left panel of Fig.~\ref{fig:Seq1}), and see a partial recovery of the spinless bosons case (Fig.~\ref{fig:Seq0}). Correlations that involve higher energy levels ($\sigma=0$ and $\sigma=-1$) are orders of magnitude weaker, at the considered temperatures, as shown in the right panels of Fig~.\ref{fig:Seq1}.

\begin{figure}[t]
\begin{center}
\includegraphics[width=1.0\columnwidth]{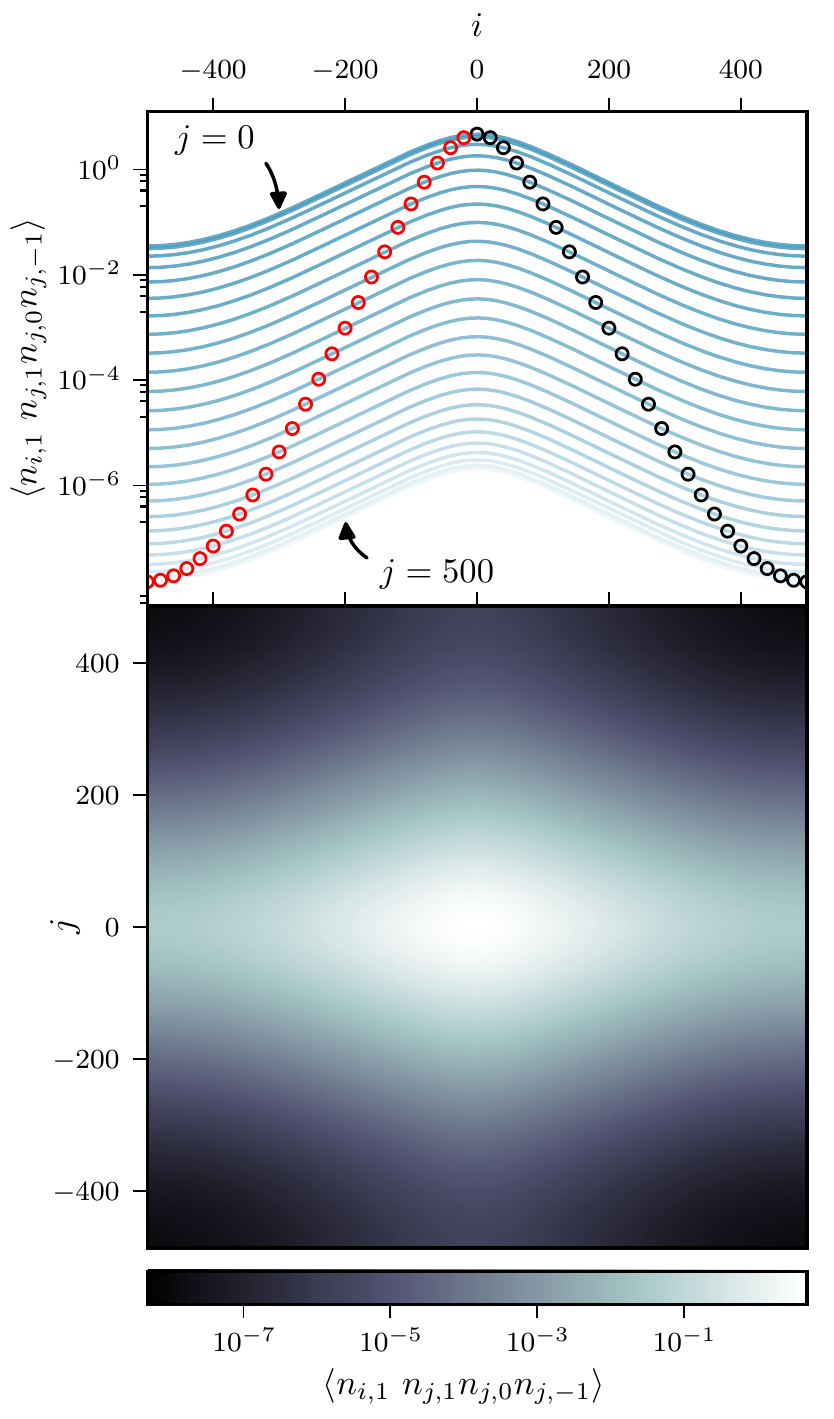}
\end{center}
\caption{Lower panel: four-level correlations $\avg{B}{N}{n_{i,1}n_{j,1}n_{j,0}n_{j,-1}}$ at $\beta=1/t$ for $N=1000$ spin-1 bosons on $L=1001$ sites with no magnetic field applied ($h=0$).  Upper panel: Horizontal cuts from the upper half of the $\avg{B}{N}{n_{i,1}n_{j,1}n_{j,0}n_{j,-1}}$ heat-map at different values of the index $j>0$. The circled data points are calculated using \Eqref{eq:corrzetaZd}, where the red-circled points are the fully degenerate case $i=-j$, while the black-circled points are $\lim_{i\to j}\avg{B}{N}{n_{i,1}n_{j,1}n_{j,0}n_{j,-1}}=\avg B N {\tbinom{n_j}{4}}$.}
\label{fig:Seq1_4points}
 \end{figure}

Finally, we turn to correlations between a set of levels that is partially degenerate, an interesting feature of spin-1 bosons.  We consider the four-level (disconnected) correlations $\avg{B}{N}{n_{i,1}n_{j,1}n_{j,0}n_{j,-1}}$ between the levels $\epsilon_{i,1}$, $\epsilon_{j,1}$, $\epsilon_{j,0}$ and $\epsilon_{j,-1}$, where, in the absence of a magnetic field, the last three levels are degenerate for any $j$.  We employ the bosonic version 
of \Eqref{eq:Exzetadeg} with $\zeta = 1$ with results shown in Fig.~\ref{fig:Seq1_4points}.
%
%%%%%%%%%%
%%%%%%%%%%
According to \Eqref{eq:njpmmkBl}, the results that we obtain, in this case, also represent
$ \avg{B}{N}{n_{i,\sigma}\tbinom{n_{j,\sigma^\prime}}{3}}$, for any $\sigma$ and $\sigma^\prime$ $\in\{1, 0, -1\}$. For the fully degenerate case $i=-j$, we use \Eqref{eq:corrzetaZd}. Once more, \Eqref{eq:njpmmkBl} guarantees that $ \avg{B}{N}{n_{i,1}n_{-i,1}n_{-i,0}n_{-i,-1}}=\avg{B}{N}{\tbinom{n_{i,\sigma}}{4}}$. Therefore, we use $\avg{B}{N}{\tbinom{n_{i,\sigma}}{4}}$ instead of $\avg{B}{N}{n_{i,\sigma}^4}$, for the diagonal elements of the 4-level disconnected correlations presented in the lower panel of Fig.~\ref{fig:Seq1_4points}. The consistency of our calculations using different equations and methods described herein is demonstrated in the upper panel in analogy with Fig.~\ref{fig:Seq0}.

% ---------------------------------------------------------------------------------
\section{Discussion}
\label{sec:V}
% ---------------------------------------------------------------------------------

\begin{table}[t!]
    \renewcommand{\arraystretch}{1.5}
    \setlength\tabcolsep{8pt}
  \begin{tabular}{@{}l@{}} 
   \toprule
   \textbf{Fermionic Results} \\
   \midrule 
   Level Correlations: Eq.~(\ref{eq:corrFgen}) \\
   \multirow{2}{\columnwidth}{\vspace{-1em}\begin{multline*}%
    \avg{F}{N}{\prod_{r=1}^{\ell}{\qty[n_{j_r}\gamma_{j_r}+(1-n_{j_r})(1-\gamma_{j_r})]}} = \\ 
   \frac{1}{\Z{F}{N}} \e{\sum_{r=1}^{\ell}\epsilon_{j_r}\gamma_{j_r}}\ZZ F {N-\sum_{r=1}^{\ell}\gamma_{j_r}} {j_1, j_2, \dots, j_\ell} 
   \end{multline*}} \\ \addlinespace[7.0em]
   Joint Probability Distribution: Eq.~(\ref{eq:pn1tolF}) \\
   \multirow{2}{\columnwidth}{\begin{equation*}%
   \p F{n_{j_1}, n_{j_2}, \dots, n_{j_\ell}} = 
   \frac{\e{\sum_{r=1}^\ell \epsilon_{j_r}n_{j_r}}}{\Z F N} \ZZ F{N-\sum_{r=1}^{\ell} n_{j_r}} {j_1, j_2, \dots, j_\ell}
   \end{equation*}}  \\ \addlinespace[4em] 
   Auxiliary Partition Function: Eq.~(\ref{eq:Z1Z1F_inv}) \\
   \multirow{2}{\columnwidth}{\begin{equation*}%
   \ZZ F N {j_1, j_2, \dots, j_\ell}=\sum_{k=0}^{N}(-1)^k\Z B k (\{j_1, j_2, \dots, j_\ell\}) \Z F {N-k}
   \end{equation*}} \\ \addlinespace[4em] 
    \toprule
   \textbf{Bosonic Results} \\
   \midrule 
   Level Correlations: Eq.~(\ref{eq:njpmmkBl}) \\
   \multirow{2}{\columnwidth}{\vspace{-1em}\begin{multline*}%
    \avg{B}{N}{\prod_{r=1}^{\ell}{\binom{n_{j_r}-q_{j_r}+m_{j_r}}{m_{j_r}}}} = \\
    \frac{1}{\Z{B}{N}} \e{\sum_{r=1}^{\ell}\epsilon_{j_r}q_{j_r}}\ZZZ B {N-\sum_{r=1}^{\ell}q_{j_r}} 
{\jdd{1}{1},\dots, \jdd{1}{m_{j_1}}, \dots \jdd{\ell}{1},\dots, \jdd{\ell}{m_{j_\ell}}}
   \end{multline*}} \\ \addlinespace[7.0em]
   Joint Probability Distribution: Eq.~(\ref{eq:pn1tolB}) \\
   \multirow{2}{\columnwidth}{\begin{equation*}%
    \p B{n_{j_1}, n_{j_2}, \dots, n_{j_\ell}}=\frac{\e{\sum_{r=1}^\ell \epsilon_{j_r}n_{j_r}}}{\Z B N} \ZZ B {N-\sum_{r=1}^{\ell} n_{j_r}} {j_1, j_2, \dots, j_\ell}
\end{equation*}} \\ \addlinespace[4em] 
   Auxiliary Partition Function: Eq.~(\ref{eq:Z1Z1B_inv}) \\
   \multirow{2}{\columnwidth}{\begin{equation*}%
           \ZZ B N {j_1, j_2, \dots, j_\ell}=\sum_{k=0}^{N}(-1)^k\Z F k (\{j_1, j_2, \dots, j_\ell\}) \Z B {N-k}
   \end{equation*}} \\ \addlinespace[4em] 
   \bottomrule
  \end{tabular}
  \caption{\label{tab:summary}A summary of the main results presented in this paper that can be utilized to determine energy level occupation numbers, correlations, and probabilities for $N$ non-interacting fermions ($F$) and bosons ($B$) with energy spectra $\epsilon_i$ with $i \in \mathcal{S} = \qty{1,2,\dots,M}$ in the canonical ensemble.  Here $n_j$ is the occupation of the $j^{th}$ level,  $\gamma_{j_r} = 0,1$ and  $0 \le q_{j_r} \le m_{j_r} \in \mathbb{Z}$. All computations rely on the introduction of auxiliary partition functions that describe a modified spectra or subset of levels connected to $\mathcal{S}$ through the removal of levels or the addition of degeneracy.}
\end{table}

In summary, we have presented a statistical theory of non-interacting identical quantum particles in the canonical ensemble,  providing a unified framework that symmetrically captures both fermionic and bosonic statistics.  Table~\ref{tab:summary} includes a listing of our most important results for fermions and bosons. We achieve this by: (1) Representing correlations (Eqs.~(\ref{eq:corrFgen}) and (\ref{eq:njpmmkBl})) and joint probability distributions ((\ref{eq:pn1tolF}) and (\ref{eq:pn1tolB})) via auxiliary partition functions. (2) Deriving general relations between the canonical partition function of a given spectrum and that of the auxiliary partition function describing a spectral subset, as captured by Eqs.~(\ref{eq:Z1Z1}), (\ref{eq:Z1Z1F_inv}) and (\ref{eq:Z1Z1B_inv}). 

These key equations can be manipulated to simplify the derivation of the known recursive relations for partition functions in the canonical ensemble and lead immediately to generalizations, and more importantly, provide useful formulas for calculating the correlations between degenerate energy levels and for calculating higher moments of the occupation numbers distribution. Also, Eqs.~\eqref{eq:corrFgen} and  \eqref{eq:njpmmkBl} can be used to reduce the complexity order of the desired correlations, or,  to relate them to the occupation numbers of the involved levels and correlations between entirely degenerate levels (see \Eqref{eq:corrzetaZd}).  Moreover, the ability to manipulate the way an auxiliary partition function is built out of other ones, allows us to construct a systematic approach towards the decomposition of many-energy level correlations in terms of individual level occupancies. This reflects the additional constraints between energy levels due to fixed $N$ even in the absence of interactions that are not present in a grand canonical description.  Thus, we present an approach to working in the canonical ensemble that includes a generalization of Wick's theorem, where we obtain previous results for non-degenerate levels \cite{Schonhammer:2017,GiraudGrabschTexier:2018} and extend them to the case of a degenerate spectrum.

Interestingly, despite the substantial difference between fermionic and bosonic statistics, the resulting formulas show evident similarity. If we compare Eqs.~(\ref{eq:njFrecursive1}), (\ref{eq:njF_ZN}) and (\ref{eq:ZNrecursiveF}).  with Eqs.~(\ref{eq:njBrecursive1}), (\ref{eq:njB_ZN}) and (\ref{eq:ZNrecursiveB}), respectively, we see that the differences between the fermionic and the bosonic formulas can be captured by simple $\pm 1$ factors.  In view of the current theory, such similarity is associated with the interplay between fermionic and bosonic auxiliary partition functions. The inverted symmetry between the two distinct statistics is apparent via a comparison of Eqs.~(\ref{eq:ZNF}) and (\ref{eq:ZNB}) with Eqs.~(\ref{eq:ZNF_inv}) and (\ref{eq:ZNB_inv}) reflective of the fact that \emph{adding} a fermionic energy level to the partition function is similar to \emph{excluding} a bosonic one and vice-versa.

The presented formulas for combining and resolving auxiliary partition functions allows for their construction via different routes which we have utilized to obtain exact expressions for the decomposition of correlations in terms of single-level occupation numbers. These different forms may also have value in overcoming the known numerical instabilities of the recursive formula for the fermionic partition function due to influence alternating signs \cite{Schonhammer:2017,SchmidtSchnack:1999}. In addition, the simplicity of the presented theory suggests a possible generalization to cover different energy-levels occupation-constraints beyond the fermionic and bosonic ones.

We envision the results presented herein could have applications in the computation of entanglement entropy in the presence of super-selection rules, as well as in modelling cold atom experiments.  In the context of quantum information, the spectrum of the reduced density matrix corresponding to a mode bipartition of a state of conserved number $N$ of itinerant particles on a lattice can be associated with that of a fictional entanglement Hamiltonian.  For non-interacting particles, the entanglement entropy can be obtained via the so-called correlation matrix method \cite{Peschel:2003,PeschelEisler:2009,EislerPeschel:2017,Peschel:2012:000000} which requires the evaluation of the canonical partition function of the resulting non-interacting entanglement Hamiltonian.  For trapped ultra-cold atoms at low densities where $N$ is fixed and interactions can be neglected, the analysis of experimental results in the physically correct canonical ensemble provides improved thermometry, especially for the case of fermions.

Finally, the ability to directly study level statistics in the canonical ensemble for bosons and fermions may have pedagogical value in the teaching of statistical mechanics, where the more physical concept of a fixed number of particles is quickly jettisoned and replaced with a grand canonical reservoir for the sake of simplifying derivations.
% ---------------------------------------------------------------------------------
\acknowledgements
% ---------------------------------------------------------------------------------
We thank D. Clougerty for bringing our attention to Ref.~[\onlinecite{DentonMuhlschlegelScalapino:1973}] and K.~Sch{\"o}nhammer for discussions at an early stage of this work. We would like to thank I.~Hamammu for pointing out the relation to nuclear physics. This research was supported in part by the National Science Foundation (NSF) under awards DMR-1553991 (A.D.) and DMR-1828489 (H.B.). All computations were performed on the Vermont Advanced Computing Core supported in part by NSF award No.~OAC-1827314.  

% ---------------------------------------------------------------------------------

% ---------------------------------------------------------------------------------
% appendix
% ---------------------------------------------------------------------------------
\appendix

\section{The derivation of \Eqref{eq:LinearB}}
\label{Appendix:A}

Starting with the sets of levels $\mathcal{S}_r=\{i_1, i_2, \dots, i_r\}\subset\mathcal{S}_\ell=\{j_1, j_2, \dots, j_\ell\}$ and using \Eqref {eq:Z1Z1B_inv} we have
\begin{equation}
\ZZZs B {N-r} {\mathcal{S}_r} = \sum_{k=0}^{\ell-r}(-1)^k\ZZs F k {\mathcal{S}_r} ( \mathcal{S}_\ell) \ZZZs B {N-r-k} {\mathcal{S}_\ell}
\label{eq:corrB1}.
\end{equation}
Next, we substitute for $\ZZs F k {\mathcal{S}_r} (\mathcal{S}_\ell)$, using Eq.~(\ref{eq:corrF2}), except for the last term  $(-1)^{\ell-r}\ZZs F {\ell-r} {\mathcal{S}_r} ( \mathcal{S}_\ell) \ZZZs B {N-\ell} {\mathcal{S}_\ell}$ which we separate from the rest of the previous summation, thus we obtain
\begin{align}
\ZZZs B {N-r} {\mathcal{S}_r} =&\sum_{k=0}^{\ell-r-1}\!\!\sum_{m=0}^{k}(-1)^{m+k}\Z B {m}(\mathcal{S}_r) \Z F {k-m}(\mathcal{S}_\ell) \ZZZs B {N-r-k} {\mathcal{S}_\ell}\nonumber\\
&+(-1)^{\ell-r}\ZZs F {\ell-r} {\mathcal{S}_r} ( \mathcal{S}_\ell) \ZZZs B {N-\ell} {\mathcal{S}_\ell}
\label{eq:corrB3-}.
\end{align}
If we rearrange the summations and perform the indexes change $k\to k-r+1$ and $m\to m-r$, we get
\begin{align}
\ZZZs B {N-r} {\mathcal{S}_r} \!\!=\!\!&\sum_{m=r}^{\ell-1}(-1)^{m+1}\Z B {m-r}(\mathcal{S}_r) \!\!\!\!\!\nonumber \\
                              &\times \sum_{k=m-1}^{\ell-2}\!\!\!\!(-1)^{k}\Z F {k-m+1}(\mathcal{S}_\ell) \ZZZs B {N-k-1} {\mathcal{S}_\ell}\nonumber\\
&+(-1)^{\ell-r}\ZZs F {\ell-r} {\mathcal{S}_r} ( \mathcal{S}_\ell) \ZZZs B {N-\ell} {\mathcal{S}_\ell}
\label{eq:corrB3}.
\end{align}
Now, using Eq.~(\ref{eq:corrB}), we substitute for $\ZZZs B {N-\ell} { \mathcal{S}_\ell}$ and $\ZZZs B {N-r} {\mathcal{S}_r}$ as well as the APF $\ZZs F {\ell-r} {\mathcal{S}_r} ( \mathcal{S}_\ell) =\e {\sum_{j_\nu\in\mathcal{S}_\ell\setminus\mathcal{S}_r}\epsilon_{j_\nu}}$. After multiplying the resulting equation by $\frac{(-1)^{r-1}\e {\sum_{i_\nu\in\mathcal{S}_r}\epsilon_{i_\nu}}}{\Z B N}$, we can write 
\begin{equation}
\Y B 0+\sum_{m=r}^{\ell-1}A_{B, m}(\mathcal{S}_r)\Y B m=b_{B}(\mathcal{S}_r)
\label{eq:LinearB1appendix}.
\end{equation}
where
\begin{align}
\Y B {1\leq m\leq\ell-1}=
\frac{(-1)^{m+1}}{\Z B N}\!\!\!\! \sum_{k=m-1}^{\ell-2}\!\!\!\! (-1)^k\Z F {k-m+1}(\mathcal{S}_\ell) \ZZZs B {N-k-1} {\mathcal{S}_\ell}
\label{eq:YmBappendix},
\end{align}
$\Y B 0=(-1)^{\ell-1}\avg B N {n_{j_1}n_{j_2}\dots n_{j_\ell}}$, $A_{0}=1$, $A_{ 0<m<r}=0$ and $A_{ r\le m\le\ell-1}(\mathcal{S}_r)=(-1)^{r-1}\e {\sum_{i_\nu\in\mathcal{S}_r}\epsilon_{i_\nu}}\Z B {m-r}(\mathcal{S}_r)$. Also, the term $b_B(\mathcal{S}_r)=(-1)^{r-1}\avg B N {n_{i_1}n_{i_2}\dots n_{i_r}}$.
%%%%%%%%%%%%%%%

\FloatBarrier

\nocite{apsrev41Control}
\bibliographystyle{apsrev4-1}
\bibliography{refs}

\end{document}